\documentclass[11pt,showpacs,preprintnumbers,superscriptaddress,amsmath,amssymb,nofootinbib]{revtex4}

\usepackage{graphicx}
\usepackage{dcolumn}
\usepackage{bm}
\usepackage{amssymb}
\usepackage{amsmath}
\usepackage{epsfig}    
\usepackage{color}    

\def\fsl#1{\setbox0=\hbox{$#1$}                 
   \dimen0=\wd0                                 
   \setbox1=\hbox{/} \dimen1=\wd1               
   \ifdim\dimen0>\dimen1                        
      \rlap{\hbox to \dimen0{\hfil/\hfil}}      
      #1                                        
   \else                                        
      \rlap{\hbox to \dimen1{\hfil$#1$\hfil}}   
      /                                         
   \fi}

\begin{document} 

\title{Probing exotic Higgs sectors from the precise measurement of Higgs boson couplings}
\preprint{UT-HET 077}
\author{Shinya Kanemura}
\affiliation{Department of Physics, University of Toyama, \\3190 Gofuku, Toyama 930-8555, Japan}
\author{Mariko Kikuchi}
\affiliation{Department of Physics, University of Toyama, \\3190 Gofuku, Toyama 930-8555, Japan}
\author{Kei Yagyu}
\affiliation{Department of Physics, National Central University, \\Chungli 32001, Taiwan}
\begin{abstract}

We study coupling constants of the standard model like Higgs boson with the gauge bosons 
$hZZ$ and $hWW$ and fermions $hf\bar{f}$ in the general Higgs sector which contains 
higher isospin representations with arbitrary hypercharge. 
In Higgs sectors with exotic Higgs representations, 
the $hZZ$ and $hWW$ coupling constants  
can be larger than those in the standard model. 
We calculate deviations in the Higgs boson couplings from standard model values in 
the model with a real or complex triplet field, the Georgi-Machacek model and 
the model with a septet scalar field. 
We also study deviations in the event rates of $h\to ZZ^*$, $h\to WW^*$, 
$h\to \gamma\gamma$, $h\to b\bar{b}$ and $h\to \tau\tau$ channels. 
\end{abstract}

\date{\today}
\maketitle

\section{Introduction}


A new boson with a mass of about 126 GeV has been discovered at the LHC~\cite{Higgs_ATLAS,Higgs_CMS}. 
The particle is likely to be the Higgs boson. 
%
%
However, this does not necessarily mean that the Higgs sector in the standard model (SM) is correct, because 
the scalar boson with the similar property to the SM Higgs boson can be predicted in 
Higgs sectors extended from the SM one. 
On the other hand, new physics models beyond the SM have often been considered by introducing extended Higgs sectors to explain 
new phenomena such as the neutrino oscillation~\cite{typeII,zee,zee-2loop,ma,krauss,aks}, 
the existence of dark matter~\cite{DM} and 
the baryon asymmetry of the Universe~\cite{ewbg-thdm}, all of which cannot be explained in the SM. 
Therefore, determining the Higgs sector is paramountly important to know 
what kind of the new physics exists at the TeV scale. 


The electroweak rho parameter is important to determine the structure of the Higgs sector. 
The experimental value of the rho parameter is close to unity~\cite{PDG}, which suggests that there 
is a global SU(2) symmetry, so-called the custodial symmetry, in the Higgs sector. 
The rho parameter strongly depends on the property of the Higgs sector; i.e., 
the number of Higgs multiplets and their hypercharges. 
In the Higgs sector composed from only SU(2) doublets and/or singlets, the rho parameter 
is unity at the tree level because of the custodial symmetry. 
Thus, these Higgs sectors can be regarded as the natural extension of the SM Higgs sector. 
On the other hand, the rho parameter deviates from unity at the tree level for
the Higgs sector with exotic representation fields such as triplets. 
In such a model, a vacuum expectation value (VEV) of such an exotic field violates 
the custodial symmetry, so that the VEV is severely constrained by the rho parameter data. 
There is another extended Higgs sector
in which an alignment of the triplet VEVs makes the rho parameter to be unity at the tree level, named as the Georgi-Machacek (GM) 
model~\cite{GM,Gunion-Vega-Wudka,Dicus,Gunion-Vega-Wudka2,AK_GM,Haber_Logan}. 
Furthermore, it is known that the addition of the isospin septet field with the hypercharge $Y=2$\footnote{Throughout the paper, we use 
the notation $Q=T_3+Y$ with $Q$ and $T_3$ to be the electromagnetic charge and third component of the isospin, respectively. } does not 
change the rho parameter from unity at the tree level. 
In order to discriminate these exotic Higgs sectors, we need to measure the other observables which are sensitive to the structure 
of the Higgs sector. 


As a striking feature of exotic Higgs sectors, there appears the $H^\pm W^\mp Z$ 
vertex at the tree level~\cite{Grifols}, where $H^\pm$ are physical singly-charged Higgs bosons. 
In the multi-doublet model, this vertex is induced at the one-loop level, so that 
the magnitude of the $H^\pm W^\mp Z$ vertex tends to be smaller than that in exotic Higgs sectors~\cite{HWZ_doublet}. 
Therefore, a precise measurement of the $H^\pm W^\mp Z$ vertex can be used to discriminate exotic Higgs sectors 
such as the GM model and the Higgs sector with a septet field. 
The feasibility of measuring this vertex has been discussed in Ref.~\cite{HWZ-LEP} at the LEPII, 
in Ref.~\cite{HWZ-Tevatron} at the Tevatron, at the LHC~\cite{HWZ-LHC} and at the International Linear Collider (ILC)~\cite{HWZ-ILC}. 
 

In this paper, 
we discuss the other method to probe or constrain exotic Higgs sectors by focusing on the 
SM-like Higgs boson couplings with the gauge bosons $hVV$ ($V=W$ and $Z$). 
At present, this approach is quite timely, because the Higgs boson like particle has already been found. 
The current accuracy of the measurement of the $hWW$ and $hZZ$ couplings at the LHC has been 
analysed to be in Ref.~\cite{Plehn1}, where the data collected in 2011 and 2012 are used.  
The Higgs boson couplings will be measured at future colliders as precisely as possible. 
For example, the $hVV$ couplings are supposed to be measured with the 
$\mathcal{O}(10)\%$ accuracy at the Hi-Luminosity LHC with the collision energy 
to be 14 TeV and the integrated luminosity to be 3000~fb$^{-1}$~\cite{Plehn2}. 
In addition, they can be measured with the 
$\mathcal{O}(1)\%$ accuracy at the ILC with the collision energy to be 
$500$ GeV and the integrated luminosity to be $500$ fb$^{-1}$~\cite{Peskin,Plehn2}.

In models with multi-doublet structure, the magnitude of the $hVV$ vertex is 
smaller than that of the corresponding SM vertices. 
On the other hand, they can be larger than the SM prediction in exotic Higgs sectors. 
Thus, measuring the $hVV$ vertex can be the other and more important tool to constrain exotic Higgs sectors in addition to 
measuring the rho parameter and $H^\pm W^\mp Z$ vertex. 
We also discuss the deviation in the Yukawa coupling $hf\bar{f}$ as well. 
 
We first derive the formula for the $hVV$ vertex in the general Higgs sector. 
We then discuss the possible deviations in the $hVV$ and $hf\bar{f}$ vertices in several concrete extended Higgs models. 
We consider the model with a real or complex triplet field, the GM model and the model with a septet field. 
In addition, we evaluate the deviation in the event rate of the signal for the $h\to WW^*$, $h\to ZZ^*$, $h\to \gamma\gamma$, 
$h\to \tau\tau$ and $h\to b\bar{b}$ in these models at the LHC.

We will find that the deviation in the $hVV$ coupling can be as large as 
$\mathcal{O}(0.1\%)$, $\mathcal{O}(30\%)$ and $\mathcal{O}(10\%)$ 
in the model with a real or complex triplet field, the GM model and the model with a septet field, respectively 
in the allowed parameter regions by the current electroweak precision data.

\section{The $hVV$ vertex}

We consider an extended Higgs sector which contains $N$ Higgs multiplets 
$\Phi_i$ ($i=1,\dots , N$) with 
the isospin $T_i$ and the hypercharge $Y_i$. 
We assume CP conservation of the Higgs sector. 
The kinetic term in the general Higgs sector is given by 
\begin{align}
\mathcal{L}_{\text{kin}}=\sum_ic_i|D_i^\mu\Phi_i|^2,
\label{cov}
\end{align}
with $c_i=1~(1/2)$ for a complex (real; i.e., $Y=0$) Higgs field. 
%
The $W$ and $Z$ boson masses are calculated as
\begin{align}
m_W^2=\frac{g^2}{2}\sum_iv_i^2[T_i(T_i+1)-Y_i^2],\quad m_Z^2=g_Z^2\sum_iv_i^2Y_i^2, \text{ with }g_Z=\frac{g}{\cos\theta_W},
\end{align}
where $v_i\equiv \sqrt{2c_i}\langle \Phi_i^0 \rangle$ is the VEV of the $i$-th Higgs field. 
The VEV $v$ ($=(\sqrt{2}G_F)^{-1/2}\simeq$ 246 GeV) can be expressed as
\begin{align}
&v^2= 2\sum_i C_i'v_i^2,\text{with }C_i' =T_i(T_i+1)-Y_i^2. 
\end{align}
The electroweak rho parameter can then be calculated at the tree level as~\cite{rho_tree}
\begin{align}
\rho_{\text{tree}}&=\frac{m_W^2}{m_Z^2\cos^2\theta_W}
=\frac{\sum_i v_i^2[T_i(T_i+1)-Y_i^2]}{2\sum_i v_i^2Y_i^2}.\label{rho}
\end{align}

The SM-like Higgs boson $h$ can be defined by 
\begin{align}
\tilde{h}_i = R_{ih}h, 
\end{align}
where $R_{ih}$ is the element of the orthogonal matrix connecting the weak eigenbasis of CP-even scalar states $\tilde{h}_i$ 
and the mass eigenbasis. 
In this notation, $\Phi_{i=h}$ should be the isospin doublet field with $Y=1/2$. 
If there is no mixing among CP-even scalar states, $\tilde{h}_h$
can be regarded as the SM-like Higgs boson $h$.

%
The $hZZ$ and $hWW$ couplings are calculated by 
\begin{align}
g_{hVV} = g_{hVV}^{\text{SM}}\times \sum_i c_{hVV}^i
= g_{hVV}^{\text{SM}}c_{hVV}
,\quad \text{with }V=W,~Z, \label{g_hVV}
\end{align}
where $g_{hVV}^{\text{SM}}$ is the $hVV$ coupling in the SM, and  
the factor $c_{hVV}^i$ is expressed by
\begin{align}
c_{hWW}^i =  \frac{\sqrt{2}v_i[T_i(T_i+1)-Y_i^2]R_{ih}}{\sqrt{\sum_j v_j^2 [T_j(T_j+1)-Y_j^2]}},\quad 
c_{hZZ}^i =  \frac{2Y_i^2 v_i R_{ih}}{\sqrt{\sum_j Y_j^2 v_j^2}}. \label{c_hVV}
\end{align}

In the general Higgs sector, the charged (neutral) 
Nambu-Goldstone (NG) bosons can be separated from physical charged Higgs bosons (CP-odd Higgs bosons) 
by using the elements of the orthogonal matrices; 
\begin{align}
w_i^\pm = R_{iG^+}G^\pm,\quad z_i^0 = R_{iG^0}G^0, \text{ with }\sum_i R_{iG^+}^2=\sum_iR_{iG^0}^2=1,\label{RG}
\end{align}
where $w_i^\pm$ ($z_i$) is the singly-charged (CP-odd) scalar state in the weak eigenbasis. 
From the NG theorem, $R_{iG^+}$ and $R_{iG^0}$ satisfy the following relations; 
\begin{align}
\frac{g}{2}\sum_i \sqrt{c_i}C_i v_i R_{iG^+}=m_W,\quad g_Z\sum_i Y_i v_i R_{iG^0}=m_Z, 
\end{align}
where  
\begin{align}
C_i =\sqrt{T_i(T_i+1)-Y_i^2+Y_i}. 
\end{align}
In the Higgs sector with one pair of a physical singly-charged Higgs boson and a physical CP-odd Higgs boson, 
the elements given in Eq.~(\ref{RG}) are expressed by
\begin{align}
R_{iG^+}=\frac{2v_i}{v}\frac{\sqrt{c_i}C_i'}{C_i},\quad 
R_{iG^0}=\frac{Y_iv_i}{\sqrt{\sum_i Y_i^2v_i^2}}.
\end{align}

The Yukawa coupling of $h$ can be simply obtained in the Higgs sector where 
only one doublet Higgs field couples to the fermions. 
In this case, the $hf\bar{f}$ coupling $g_{hff}$ is expressed as 
\begin{align}
g_{hff} = g_{hff}^{\text{SM}}\times c_{hff},\text{ with}~c_{hff}=\frac{v}{v_i}R_{ih}. \label{c_hff}
\end{align}
In the model with multi-doublets, a discrete symmetry is necessary to realize such a situation as we discuss in the next section for 
the two Higgs doublet model (THDM).  

\section{Examples}

\begin{table}[t]
\begin{center}
{\renewcommand\arraystretch{1.3}
\begin{tabular}{|c||c|c|c|c|}\hline
Model & $\tan\beta$ &$\tan\beta'$& $c_{hWW}$ & $c_{hZZ}$  \\\hline\hline
$\phi_1+\phi_2$ (THDM) &$v_{\phi_2}/v_{\phi_1}$&$v_{\phi_2}/v_{\phi_1}$ &$\sin(\beta-\alpha)$ & $\sin(\beta-\alpha)$   \\\hline
$\phi+\chi$ (cHTM) &$\sqrt{2}v_\chi/v_\phi$&$2v_\chi/v_\phi$& $\cos\beta \cos\alpha + \sqrt{2}\sin\beta\sin\alpha$ & $\cos\beta' \cos\alpha + 2\sin\beta'\sin\alpha$ \\\hline
$\phi+\xi$ (rHTM) &$2v_\xi/v_\phi$&-& $\cos\beta \cos\alpha + 2\sin\beta\sin\alpha$ & $\cos\alpha$  \\\hline
$\phi+\chi+\xi$ (GM model) &$2\sqrt{2}v_\Delta/v_\phi$& $2\sqrt{2}v_\Delta/v_\phi$ &$\cos\beta \cos\alpha +\frac{2\sqrt{6}}{3}\sin\beta \sin\alpha$ &$\cos\beta \cos\alpha +\frac{2\sqrt{6}}{3}\sin\beta \sin\alpha$ \\\hline
$\phi+\varphi_7$ &$4v_{\varphi_7}/v_\phi$& $4v_{\varphi_7}/v_\phi$ &$\cos\beta \cos\alpha +4\sin\beta \sin\alpha$ &$\cos\beta \cos\alpha +4\sin\beta \sin\alpha$ \\\hline
\end{tabular}}
\caption{The deviations in the Higgs boson couplings from the SM values in various extended Higgs sectors. 
$\phi$, $\chi$, $\xi$ and $\varphi_7$ are respectively denoted as Higgs fields with ($T,Y$)=($1/2,1/2$), ($1,1$), ($1,0$) and ($3,2$). 
In the second and third column, $v_X$ is the VEV of the Higgs field $X$, and $v_\Delta$ is defined in Eq.~(\ref{VEV_align}). 
The mixing angle $\alpha$ is defined for each extended Higgs sector in the main text. 
}
\label{models}
\end{center}
\end{table}

We discuss the Higgs boson couplings in several concrete Higgs sectors. 
As examples, we consider the THDM, the model with a complex triplet field (cHTM), that with a real 
triplet field (rHTM), the GM model~\cite{GM} and the model with a septet Higgs filed. 
The Higgs fields content in each model is listed in Table~\ref{models}, 
where $\phi$ ($\phi_1$ and $\phi_2$ have the same quantum number as $\phi$), 
$\chi$, $\xi$ and $\varphi_7$ are respectively denoted Higgs fields with ($T,Y$)=($1/2,1/2$), ($1,1$), ($1,0$)
and ($3,2$).   
In this table, $\beta$ ($\beta'$) is the mixing angle which separates the charged (CP-odd) NG boson from 
the physical singly-charged (CP-odd) Higgs bosons. 
The mixing angle among CP-even scalar states are expressed as $\alpha$, whose definitions 
are given in the following subsequent paragraphs in each extended Higgs sector.

First, in the THDM,  
$c_{hVV}$ is calculated as $\sin(\beta-\alpha)$, where $R_{1h}=-\sin\alpha$ and 
$R_{2h}=\cos\alpha$. 
Thus, the $hVV$ vertex cannot be larger than that in the SM. 
The Yukawa couplings of $h$ depends on the variation of the THDMs. 
When a softly-broken $Z_2$ symmetry 
is imposed to the model in order to avoid the tree level flavour changing neutral current, 
there are four types of the Yukawa couplings depending on the charge assignment of the $Z_2$ charge~\cite{type_THDM}. 
The expression of the Yukawa couplings in each type is given in Ref.~\cite{AKTY}. 
In the following four extended Higgs sectors: the cHTM, the rHTM, the GM model and the model with $\varphi_7$, the 
deviation in the Yukawa coupling can be expressed by $c_{hff}=\cos\alpha/\cos\beta$. 

Second, in both the cHTM and the rHTM, 
$c_{hWW}$ can be larger than unity as listed in Table~\ref{models}, 
where the mixing angle $\alpha$ is defined by $R_{1h}=\cos\alpha$ and 
$R_{2h}=\sin\alpha$. 
Because there is no additional CP-odd scalar state in the rHTM, the mixing angle $\beta'$ cannot be defined, so that 
$c_{hZZ}$ is smaller than 1 by non-zero values of the mixing angle $\alpha$. 
In the cHTM, $c_{hZZ}$ can also be larger than 1, but the pattern of the deviation is different from $c_{hWW}$. 
In both two models, the rho parameter deviates from unity because of the non-zero VEV of the triplet field. 
The magnitude of the VEV of the complex (real) triplet field $v_\chi$ ($v_\xi$) 
is constrained to be less than about 8 GeV in the cHTM (about 6 GeV in the rHTM) from 
the experimental value of the rho parameter $\rho^{\text{exp}}=1.0008^{+0.0017}_{-0.0007}$~\cite{PDG}. 

\begin{figure}[t]
\begin{center}
\includegraphics[width=80mm]{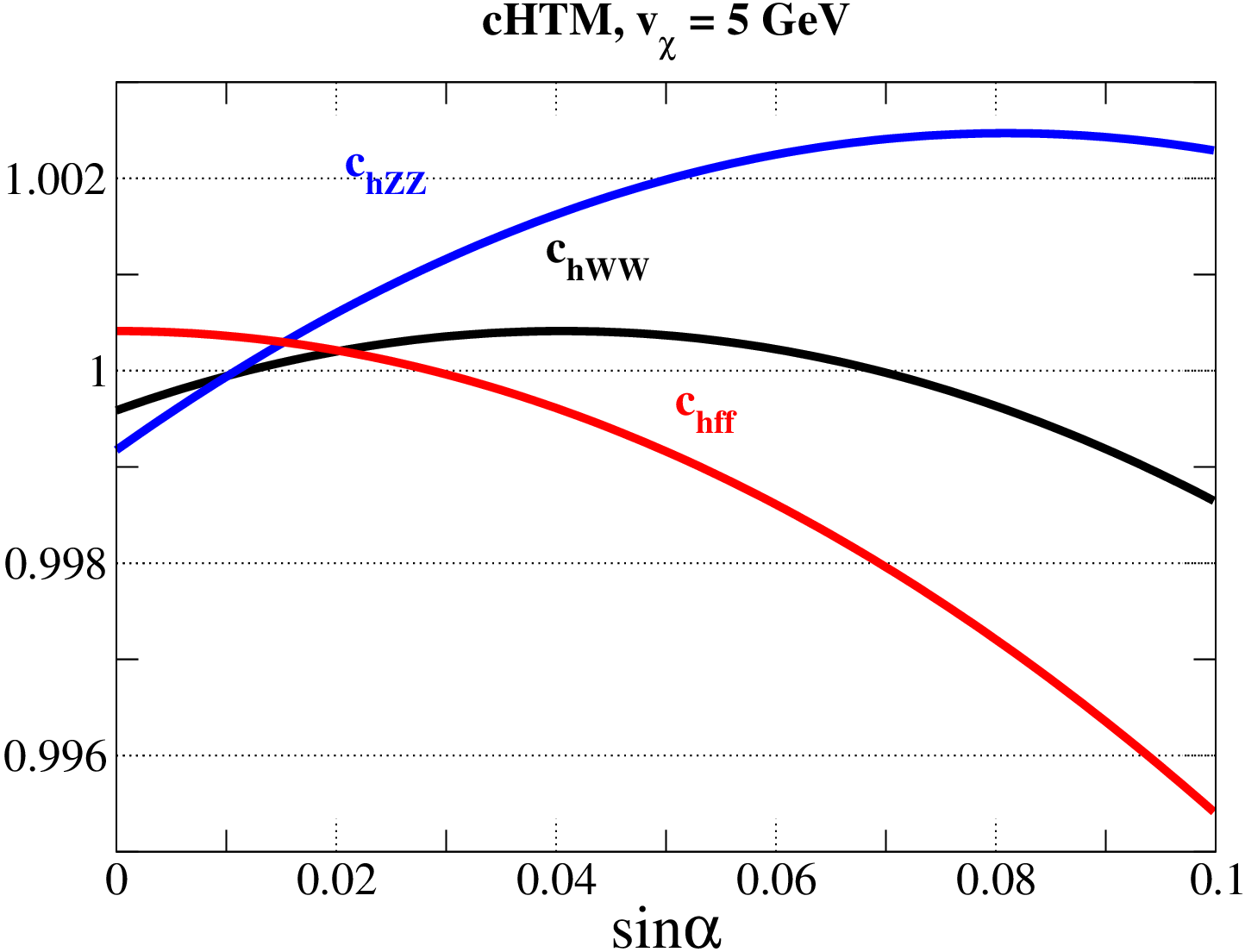}\hspace{3mm}
\includegraphics[width=80mm]{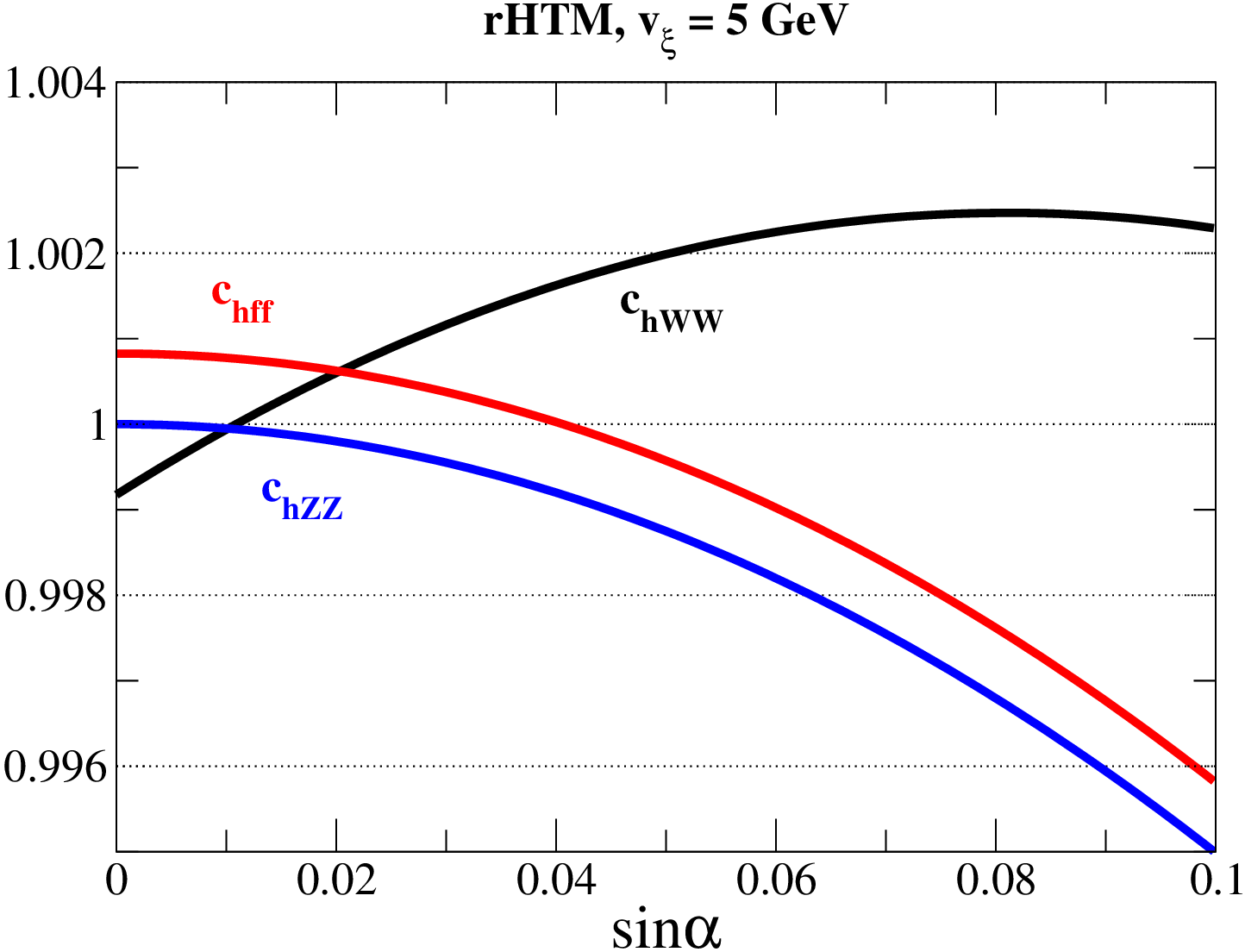}
\caption{$c_{hWW}$, $c_{hZZ}$ and $c_{hff}$ as a function of $\sin\alpha$ 
in the cHTM (left panel) and in the rHTM (right panel) for the case of $v_\chi =v_\xi= 5$ GeV. }
\label{FIG:hVV}
\end{center}
\end{figure}

In Fig.~\ref{FIG:hVV},  the deviations in the $hWW$, $hZZ$ and $hf\bar{f}$ couplings are shown as a function of $\sin\alpha$ 
in the cHTM (left panel) and the rHTM (right panel) for $v_\chi=v_\xi=5$ GeV.  
It is seen that there are regions where both $c_{hWW}$ and $c_{hZZ}$ are larger than 1 in the cHTM\footnote{Even when effects of 
the radiative corrections to the $hVV$ coupling are taken into account, the results of $c_{hVV}>1$ can be obtained~\cite{AKKY}. }, 
while only $c_{hWW}$ can be larger than 1 in the rHTM as mentioned in the above. 
The maximal allowed values for $c_{hWW}$ and $c_{hZZ}$ in the cHTM can be estimated in the case with 
$\sin\beta\ll 1$, $\sin\beta' \ll 1$ and $\sin\alpha \ll 1$ by 
\begin{align}
c_{hWW}
&=-\frac{1}{2}\left(\alpha-\sqrt{2}\beta\right)^2+1+\frac{\beta^2}{2} +\mathcal{O}(\beta^2\alpha^2),\label{hWW_ap}
\\
c_{hZZ}
&=-\frac{1}{2}\left(\alpha-2\beta^\prime\right)^2+1+\frac{3\beta^{\prime2}}{2} +\mathcal{O}(\beta^{\prime2}\alpha^2). \label{hZZ_ap}
\end{align}
From above the equations, it can be seen that the $hWW$ ($hZZ$) coupling can be taken to be a maximal value 
in the case of $\alpha = \sqrt{2}\beta$ ($\alpha =  2\beta'$). 
When we take $v_\chi =5$ GeV, 
we obtain $c_{hWW}-1\simeq 4.2\times 10^{-4}$ for $\alpha= \sqrt{2}\beta$ 
and $c_{hZZ}-1\simeq 2.5\times 10^{-3}$ for $\alpha= 2\beta^\prime$, which are consistent with the result 
in Fig.~\ref{FIG:hVV}.

Third, we discuss the GM model where a complex triplet and a real triplet fields are contained in addition to the 
doublet field. 
The doublet Higgs field and the two triplet fields can be respectively represented 
by the $2\times 2$ and $3\times 3$ matrix forms which are transformed 
under the global $SU(2)_L\times SU(2)_R$ symmetry as 
\begin{align}
\phi=\left(
\begin{array}{cc}
\phi^{0*} & \phi^+ \\
\phi^- & \phi^0
\end{array}\right),\quad 
\Delta=\left(
\begin{array}{ccc}
\chi^{0*} & \xi^+ & \chi^{++} \\
\chi^- & \xi^0 & \chi^{+} \\
\chi^{--} & \xi^- & \chi^{0} 
\end{array}\right).
\end{align}
In this model, there are three CP-even scalar states, so that there are three mixing angles which diagonalize the 
mass matrix for the CP-even scalar states in general. 
However, when the Higgs potential is constructed under the custodial $SU(2)_V$ symmetric way, 
the mass matrix for the CP-even states can be diagonalized by the single mixing angle $\alpha$ as 
\begin{align}
\left(
\begin{array}{c}
h_\phi \\
h_\xi \\
h_\chi 
\end{array}\right) =
\left(
\begin{array}{ccc}
1 & 0 & 0 \\
0 & \frac{1}{\sqrt{3}} & -\sqrt{\frac{2}{3}} \\
0 & \sqrt{\frac{2}{3}} & \frac{1}{\sqrt{3}}
\end{array}\right)
\left(
\begin{array}{ccc}
\cos\alpha & -\sin\alpha & 0 \\
\sin\alpha & \cos\alpha & 0 \\
0 & 0 & 1
\end{array}\right)
\left(
\begin{array}{c}
h \\
H_1 \\
H_5 
\end{array}\right), \label{alpha_GM}
\end{align}
where $H_5$ ($H_1$) is the neutral component of the $SU(2)_V$ five-plet (singlet) Higgs boson, and 
$h_\phi=\sqrt{2}\text{Re}(\phi^0)$, $h_\xi=\xi^0$ and $h_\chi=\sqrt{2}\text{Re}(\chi^0)$.  

When we assume that the two 
triplet VEVs are aligned by 
\begin{align}
v_\Delta^2\equiv v_\chi^2=v_\xi^2,  \label{VEV_align}
\end{align}
with $v_\chi = \langle \chi^0\rangle$ and $v_\xi = \langle \xi^0\rangle$, 
the $SU(2)_L\times SU(2)_R$ symmetry reduces to the custodial $SU(2)_V$ symmetry. 
Therefore, the rho parameter is predicted to be unity at the tree level, whose value does not depend on the magnitude of $v_\Delta$ 
as long as we assume the alignment for the triplet VEVs. 
The expressions for $c_{hVV}$ (both $c_{hWW}$ and $c_{hZZ}$ are the same in this model) are listed in Table~\ref{models}.

At the one-loop level, the modified $hVV$ couplings and existence of extra Higgs bosons can affect the 
oblique corrections to the gauge bosons, namely the Peskin-Takeuchi 
$S$, $T$ and $U$ parameters~\cite{Peskin_Takeuchi}. 
They can be expressed as
\begin{align}
S&=\frac{4s_W^2c_W^2}{\alpha_{\text{em}}}\left[\frac{\Pi^{\text{1PI}}_{\gamma\gamma}(m_Z^2)-\Pi^{\text{1PI}}_{\gamma\gamma}(0)}{m_Z^2}
+\frac{c_W^2-s_W^2}{c_Ws_W}\frac{\Pi^{\text{1PI}}_{Z\gamma}(m_Z^2)-\Pi^{\text{1PI}}_{Z\gamma}(0)}{m_Z^2}
-\frac{\Pi^{\text{1PI}}_{ZZ}(m_Z^2)-\Pi^{\text{1PI}}_{ZZ}(0)}{m_Z^2}
\right],\\
T&=\frac{1}{\alpha_{\text{em}}}\left[\frac{\Pi_{ZZ}^{\text{1PI}}(0)}{m_Z^2}-\frac{\Pi_{WW}^{\text{1PI}}(0)}{m_W^2}+\frac{2s_W}{c_W}\frac{\Pi_{Z\gamma}^{\text{1PI}}(0)}{m_Z^2}+\frac{s_W^2}{c_W^2}\frac{\Pi_{\gamma\gamma}^{\text{1PI}}(0)}{m_Z^2}\right]
+\delta T
,\label{Tpara}\\
U&=\frac{4s_W^2}{\alpha_{\text{em}}} \Big[s_W^2\frac{\Pi^{\text{1PI}}_{\gamma\gamma}(m_Z^2)-\Pi^{\text{1PI}}_{\gamma\gamma}(0)}{m_Z^2}
+2s_Wc_W\frac{\Pi^{\text{1PI}}_{Z\gamma}(m_Z^2)-\Pi^{\text{1PI}}_{Z\gamma}(0)}{m_Z^2}
+c_W^2\frac{\Pi^{\text{1PI}}_{ZZ}(m_Z^2)-\Pi^{\text{1PI}}_{ZZ}(0)}{m_Z^2}\notag\\
&-\frac{\Pi^{\text{1PI}}_{WW}(m_W^2)-\Pi_{WW}^{\text{1PI}}(0)}{m_W^2}\Big], 
\end{align}
where $\Pi_{XY}^{\text{1PI}}(p^2)$ are the 1PI diagrams for the gauge boson two point functions at the one-loop level, whose 
analytic expressions are given in Appendix~B. 
In Eq.~(\ref{Tpara}), $\delta T$ is the counter term for the $T$ parameter, which 
does not appear in models with the custodial symmetry in the kinetic term
without imposing any alignment for VEVs as in the multi-doublet models as well as the model with the septet Higgs field. 
On the other hand, in the GM model, 
we need an alignment for the triplet VEVs to maintain the custodial symmetry at the tree level as in Eq.~(\ref{VEV_align}). 
Thus, there appear contributions to the violation of the alignment at the one-loop level, in which the ultra-violet divergence 
are contained as it has already been pointed out in Ref~\cite{Gunion-Vega-Wudka2}. 
Therefore, the counter term $\delta T$ exists associated with the parameter which gives the violation of the VEV alignment, and 
it can absorb the divergence by imposing an additional renormalization condition. 
In Ref.~\cite{Englert}, $T=0$ is imposed by using this additional renormalization condition, and 
we apply the same condition in our analysis.

The experimental values for $S$ and $ T$ parameters by fixing $ U=0$ are given as~\cite{ST_126}
\begin{align}
S=0.05\pm 0.09,\quad  T=0.08\pm0.07, \label{ST_exp}
\end{align}
where the correlation coefficient is +0.91, and the reference value of the mass of the SM Higgs boson is set to be 126 GeV. 
If we further fix $T=0$, then the 95\% confidence level region for $S$ is given by $-0.11<S<0.019$.

\begin{figure}[t]
\begin{center}
\includegraphics[width=100mm]{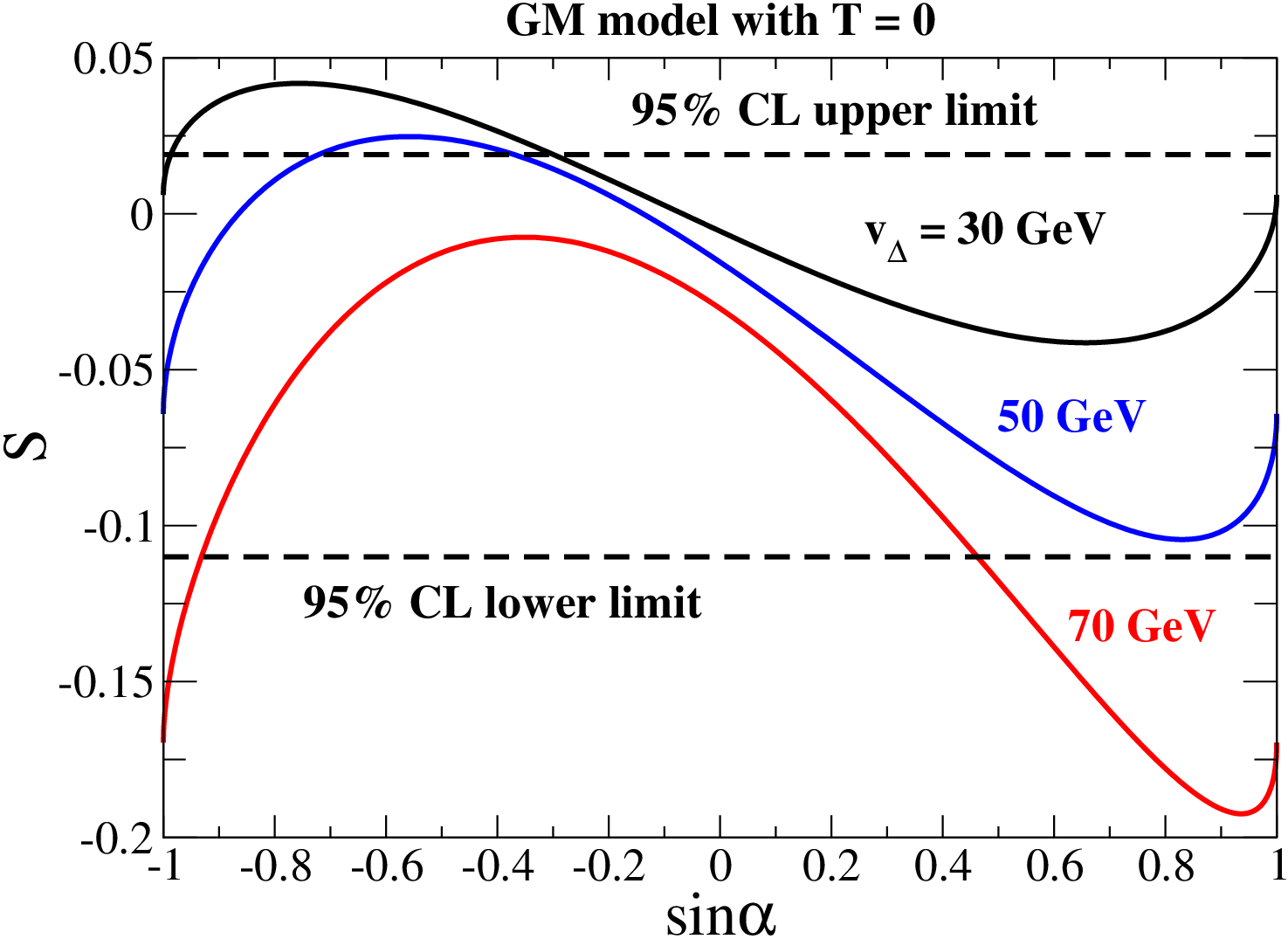}
\caption{The value for $S$ as a function of $\sin\alpha$ in the GM model for the cases with $v_\Delta =30$, 50 and 70 GeV. 
The upper and lower limits at the 95\% confidence level for the $S$ parameter are shown by the dashed curves. }
\label{S_GM}
\end{center}
\end{figure}

In Fig.~\ref{S_GM}, the $S$ parameter is shown as a function of $\sin\alpha$ which is defined in Eq.~(\ref{alpha_GM}) for the cases 
with $v_\Delta = 30$ GeV, 50 GeV and 70 GeV. 
All the masses of the extra Higgs bosons are taken to be 500 GeV in this analysis. 
When $v_\Delta$ is taken to be 30 GeV and 50 GeV (70 GeV), the ranges of $-0.99<\sin\alpha<-0.31$ and $-0.72<\sin\alpha<-0.38$ 
($-0.93<\sin\alpha<0.46$) are excluded (allowed) by the $S$ parameter at the 95\% confidence level.

\begin{figure}[t]
\begin{center}
\includegraphics[width=100mm]{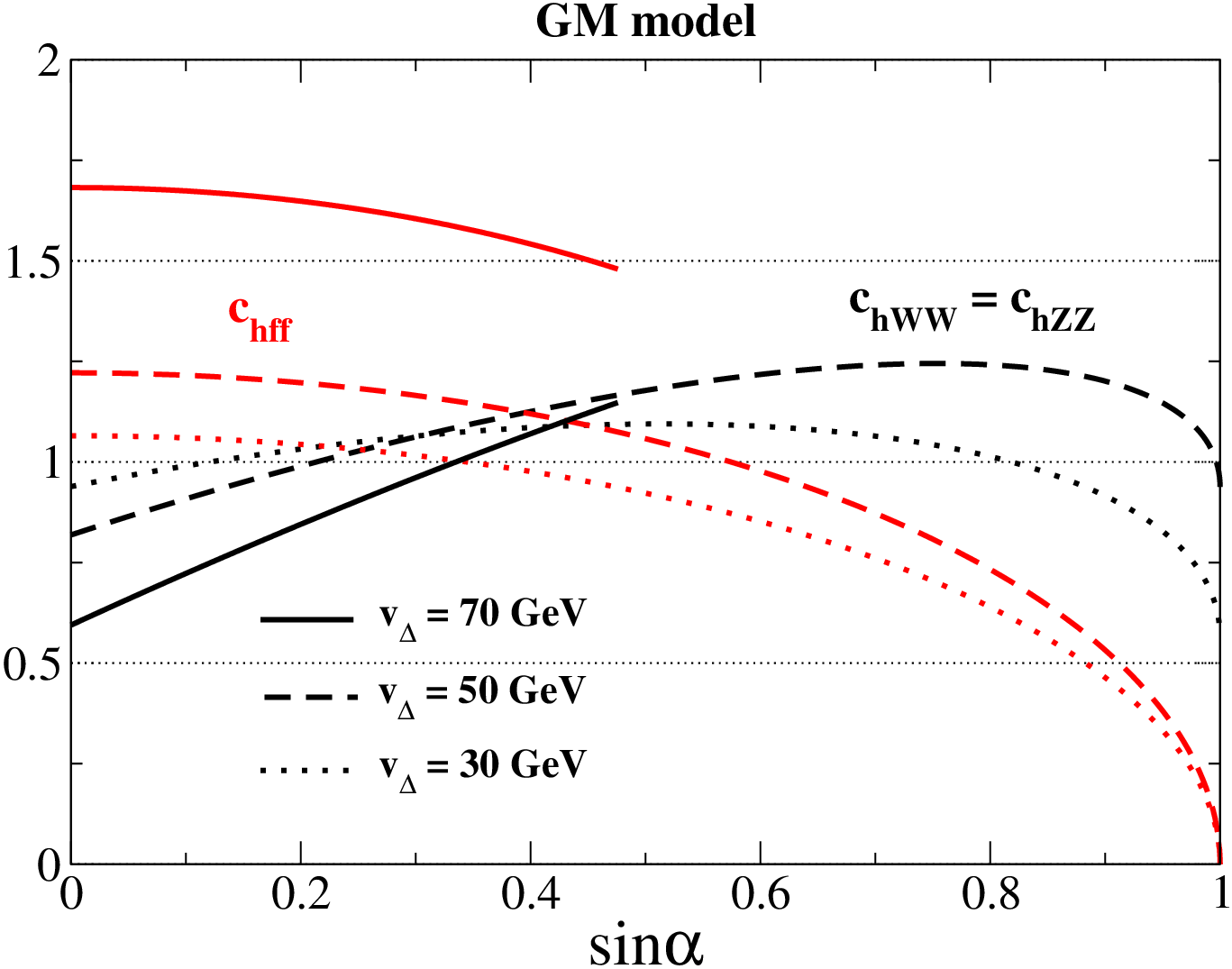}
\caption{$c_{hWW}$, $c_{hZZ}$ and $c_{hff}$ as a function of $\sin\alpha$ 
in the GM model for the cases with $v_\Delta =30$, 50 and 70 GeV. }
\label{FIG:hVV_GM}
\end{center}
\end{figure}

In Fig.~\ref{FIG:hVV_GM}, the deviations in the $hWW$, $hZZ$ and $hf\bar{f}$ couplings are shown as a function of $\sin\alpha$ 
in the GM model for the cases with $v_\Delta=30$, 50 and 70 GeV.  
The deviations in $hWW$ and $hZZ$ couplings are the same. 
The allowed maximal values for $c_{hWW}~(=c_{hZZ})$ are about 1.1, 1.3 and 1.2 for the cases of $v_\Delta=30$, 50 and 70 GeV, respectively. 
When $c_{hWW}$ and $c_{hZZ}$ are getting the maximal values, $c_{hff}$ is smaller than 1.

\begin{figure}[t]
\begin{center}
\includegraphics[width=100mm]{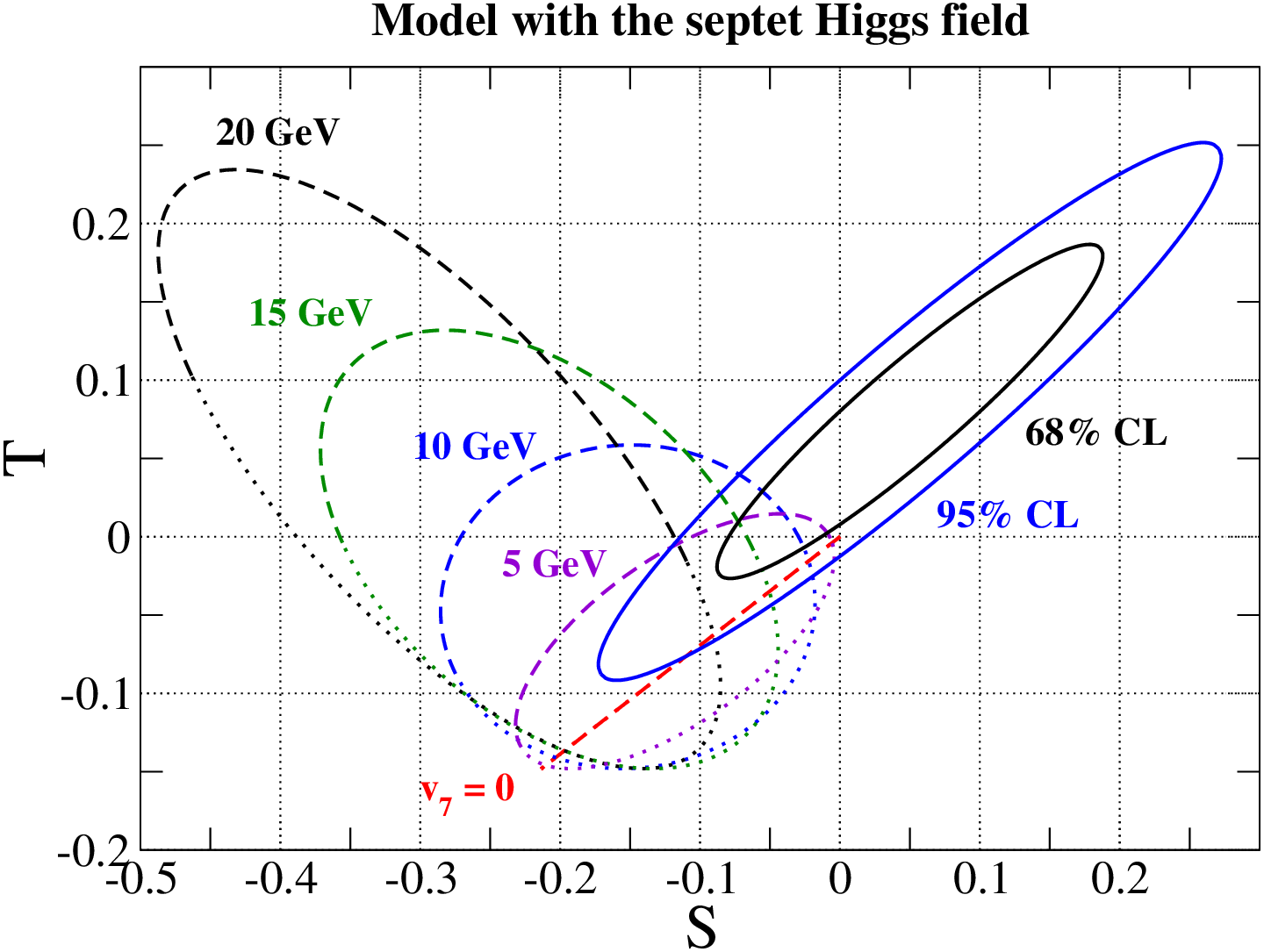}
\caption{Prediction of the $S$ and $T$ parameters in the model with the septet Higgs field. 
The regions within the black (blue) solid ellipse are allowed at the 68\% (95\%) confidence level.
Each dashed (dotted) curve shows the results in the cases with $v_7=0$, 5, 10, 15 and 20 GeV, 
where the value of $\sin\alpha$ is taken to be from 0 to 1 (from 0 to $-1$).
The left and right points where the dashed and dotted curves are crossing correspond to the case with $\sin\alpha=0$ and 
$\sin\alpha=\pm1$. 
When $\sin\alpha$ shifts from 0 to positive (negative) values, the predictions are moved to the upper (lower) region along  
the dashed (dotted) curves. 
}
\label{ST_7plet}
\end{center}
\end{figure}

\begin{figure}[t]
\begin{center}
\includegraphics[width=100mm]{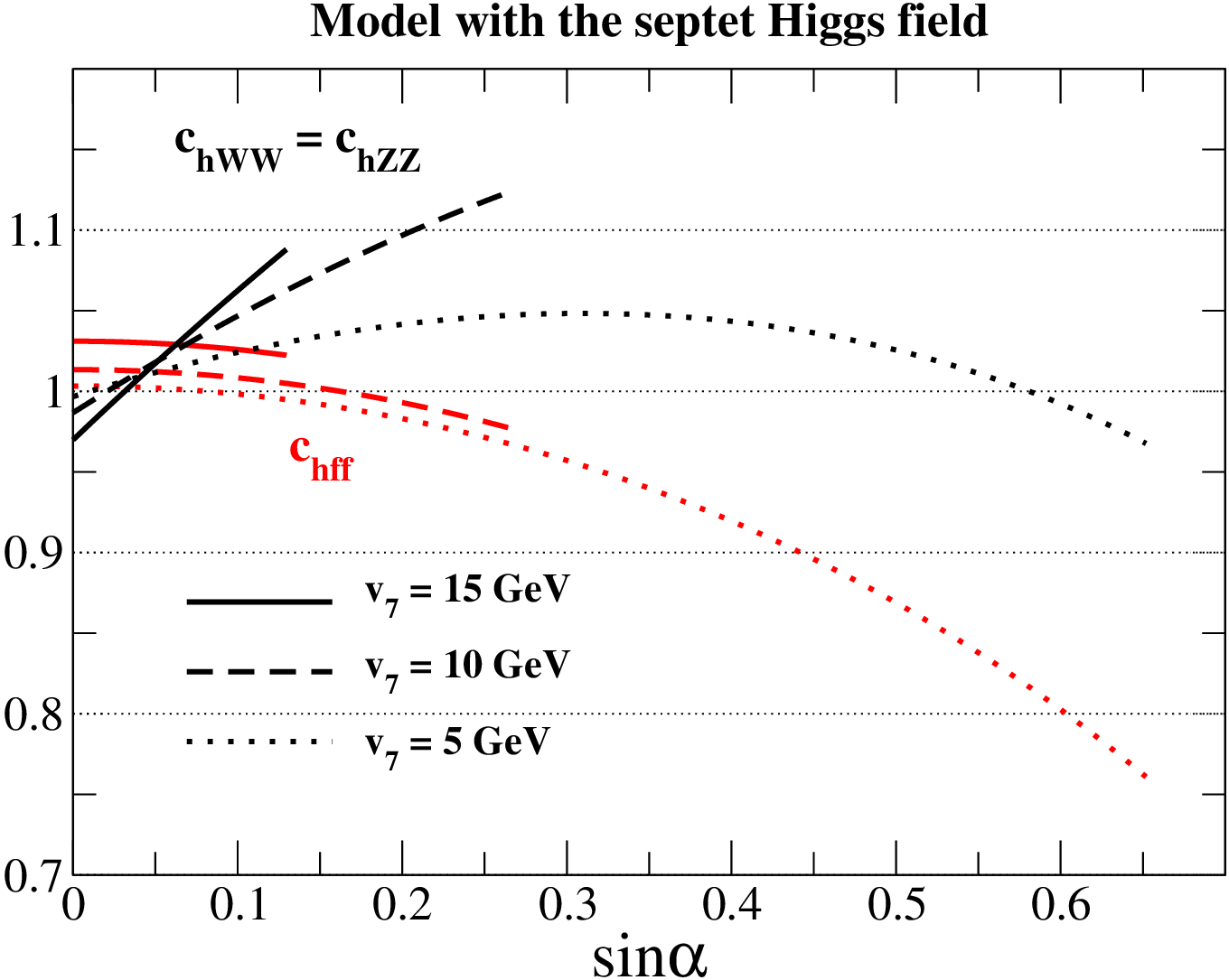}
\caption{$c_{hWW}$, $c_{hZZ}$ and $c_{hff}$ as a function of $\sin\alpha$ 
in the model with the septet Higgs field $\varphi_7$ for the cases with $v_7 =30$ and 50 GeV. }
\label{FIG:hVV_7}
\end{center}
\end{figure}

Finally, we discuss the model with the septet Higgs field $\varphi_7$ with $Y=2$. 
The expressions for $c_{hVV}$ are listed in Table~\ref{models}. 
Similar to the GM model, both $c_{hWW}$ and $c_{hZZ}$ are coincide with each other. 
Detailed calculations of the Higgs potential in this model are given in Appendix~A. 
In the GM model, although we need the alignment for the VEVs of the two triplet fields in order to keep $\rho_{\text{tree}}=1$, 
it can be directly confirmed that in the model with $\varphi_7$, the VEV of $\varphi_7$ does not change $\rho_{\text{tree}}=1$ 
from Eq.~(\ref{rho}).  
However, non-zero values for $v_7$ and the mixing angle between the CP-even Higgs bosons $\alpha$ 
which is defined in Eq.~(\ref{alpha_7}) can be constrained by the $S$ and $T$ parameters. 
Unlike the GM model, this model respects the custodial symmetry without any alignment for VEVs, 
the counter term $\delta T$ in the $T$ parameter does not exist. 
Thus, constraints from the $S$ and $T$ parameters can be applied to this model in the same way as in the SM. 
The analytic expressions for the $\Pi_{XY}^{\text{1PI}}(p^2)$ functions are given in Appendix~B.

In Fig.~\ref{ST_7plet}, the predictions of the $S$ and $T$ parameters are plotted in the cases with several fixed values of $v_7$ 
in the model with the septet Higgs field. 
All the masses of the extra Higgs bosons are taken to be 500 GeV. 
The dashed (dotted) curves show the prediction when the value of $\sin\alpha$ is changed from 0 to $1$ (0 to $-1$). 
It can be seen that the case with $v_7>20$ GeV is highly constrained by the $S$ parameter. 
The allowed maximal (minimal) values for $\sin\alpha$ at the 95\% confidence level can be obtained as $0.73(-0.73)$, 0.65($-0.098$), 0.27($-0.11$), 0.13$(-0.13)$ and 0.042($-0.15$) 
in the cases with $v_7=0$, 5 GeV, 10 GeV, 15 GeV and 20 GeV, respectively.


In Fig.~\ref{FIG:hVV_7}, we show $c_{hVV}$ and $c_{hff}$ as a function of $\sin\alpha$ 
in the model with $\varphi_7$ for the cases of $v_7 = 5$, 10 GeV and 15 GeV. 
The value of $\sin\alpha$ is taken to be positive. 
Only the parameter regions allowed by the $S$ and $T$ parameters at the 95\% confidence level are shown in this plot.
The allowed maximal values of $c_{hVV}$ is about 1.05, 1.12 and 1.09 in the cases with $v_7=5$ GeV, 10 GeV and 15 GeV, respectively.

\section{The event rates}

The production cross section and decay branching fraction of the SM-like Higgs boson $h$ 
can be modified by the deviation in $c_{hWW}$, $c_{hZZ}$ and $c_{hff}$ from unity. 
In order to clarify the deviations in the event rates of various modes for $h$ from the SM prediction, 
we define the ratio of the event rate: 
\begin{align}
R_X = \frac{\sigma_{h}\times \text{BR}(h\to X)}{\sigma_{h}^{\text{SM}}\times \text{BR}(h\to X)^{\text{SM}}}, 
\end{align}
where $\sigma_{h}^{\text{SM}}$ and $\text{BR}(h\to X)^{\text{SM}}$ [$\sigma_{h}$ and $\text{BR}(h\to X)$] 
are the production cross section of $h$ and the branching fraction of the $h\to X$ decay mode in the SM [in extended Higgs sectors]. 
So far, the Higgs boson search has been done in the following five channels at the LHC~\cite{Higgs_ATLAS,Higgs_CMS}; 
$pp\to h \to \gamma\gamma$, $pp\to h\to ZZ^{*}$, $pp\to h\to WW^*$, $pp\to h\to \tau\tau$ and $q\bar{q}' \to Vh \to Vb\bar{b}$, 
where $pp\to h$ is the inclusive Higgs boson production and $q\bar{q}' \to Vh$ is the weak vector boson associated production.  
The inclusive production cross section is almost determined by the gluon fusion production process: $gg\to h$, so that 
the modified one can be approximately expressed by $\sigma_h^{\text{SM}}(gg\to h)\times c_{hff}^2$. 
The cross section for the gauge boson associate production process can be changed by $\sigma_h^{\text{SM}}(q\bar{q}'\to Vh)\times c_{hVV}^2$. 

In general, there are charged Higgs bosons in extended Higgs sectors, and they
can contribute to the $h\to\gamma\gamma$ mode in addition to the 
W boson and top quark loop contributions. 
In the following analysis, we ignore these contributions in order to focus on the effects of 
the modified $hWW$ and $hf\bar{f}$ couplings to the $h\to \gamma\gamma$ mode. 

\begin{figure}[t]
\begin{center}
\includegraphics[width=75mm]{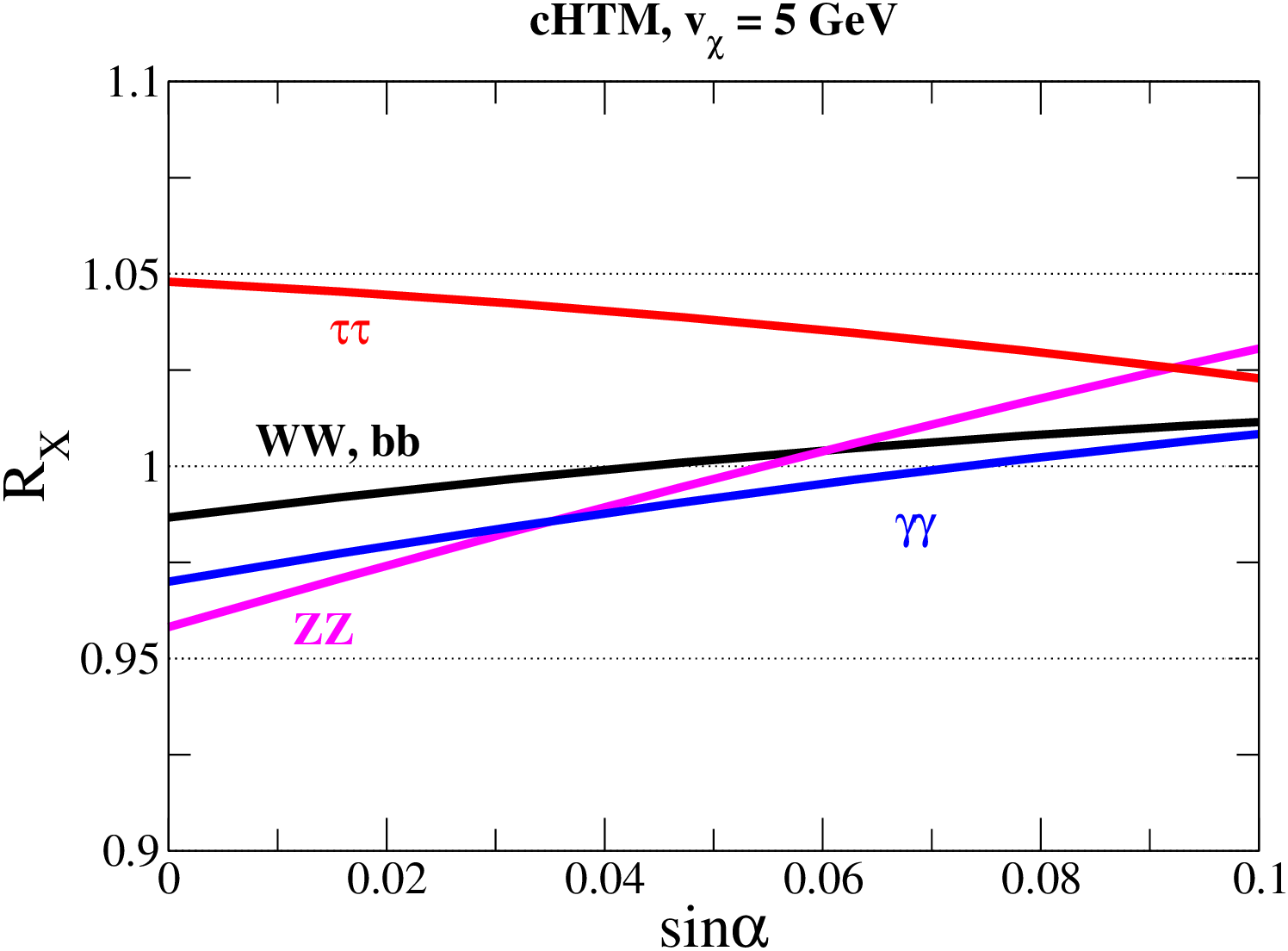}\hspace{5mm}
\includegraphics[width=75mm]{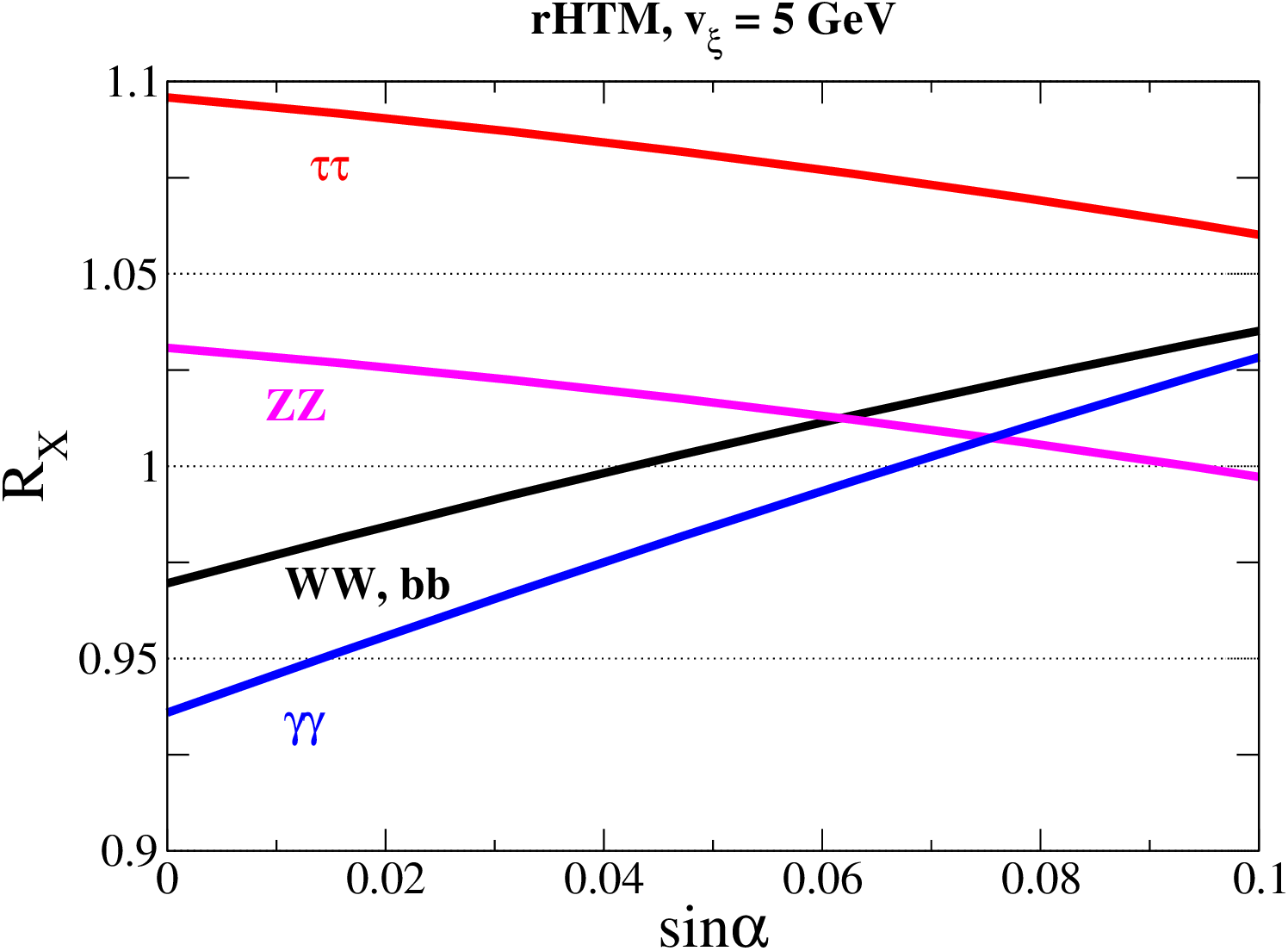}
\caption{$R_{X}$ for the various modes as a function of $\sin\alpha$ in the cHTM (left panel) and 
in the rHTM (right panel) for the case of $v_\chi=v_\xi=5$ GeV. }
\label{FIG:R1}
\end{center}
\end{figure}

In Fig.~\ref{FIG:R1}, the $R_X$ values are plotted as a function of $\sin\alpha$ in the cHTM (left panel) and 
in the rHTM (right panel) for the case of $v_\chi=v_\xi=5$ GeV. 
In both the cHTM and the rHTM, $R_{\tau\tau}$ is larger than 1 in the region of $0\leq\sin\alpha\leq 0.1$, 
because both the production cross section and the decay rate of $h\to \tau\tau$ are enhanced by the factor $c_{cff}^2\simeq 1/\cos^2\beta$ 
for the case of $\sin\alpha\sim 0$. 
When $\sin\alpha$ is taken to be larger values, $R_{\tau\tau}$ is getting smaller due to the suppression of $\cos^4\alpha$, while  
the other $R_X$ values monotonically increase in the cHTM. 
In the rHTM, $R_{ZZ}$ shows a similar behavior to that of $R_{\tau\tau}$.

\begin{figure}[t]
\begin{center}
\includegraphics[width=50mm]{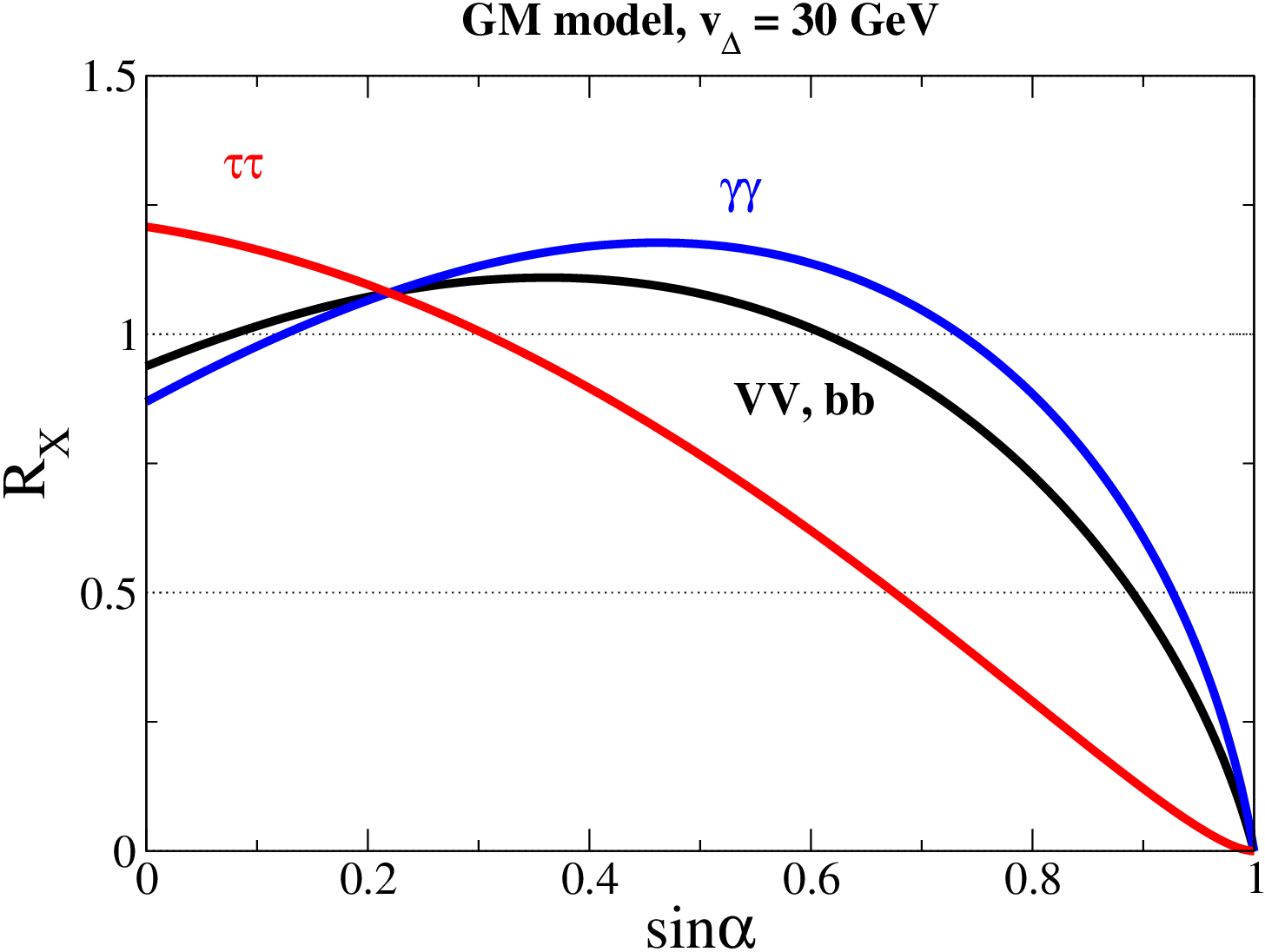}\hspace{3mm}
\includegraphics[width=50mm]{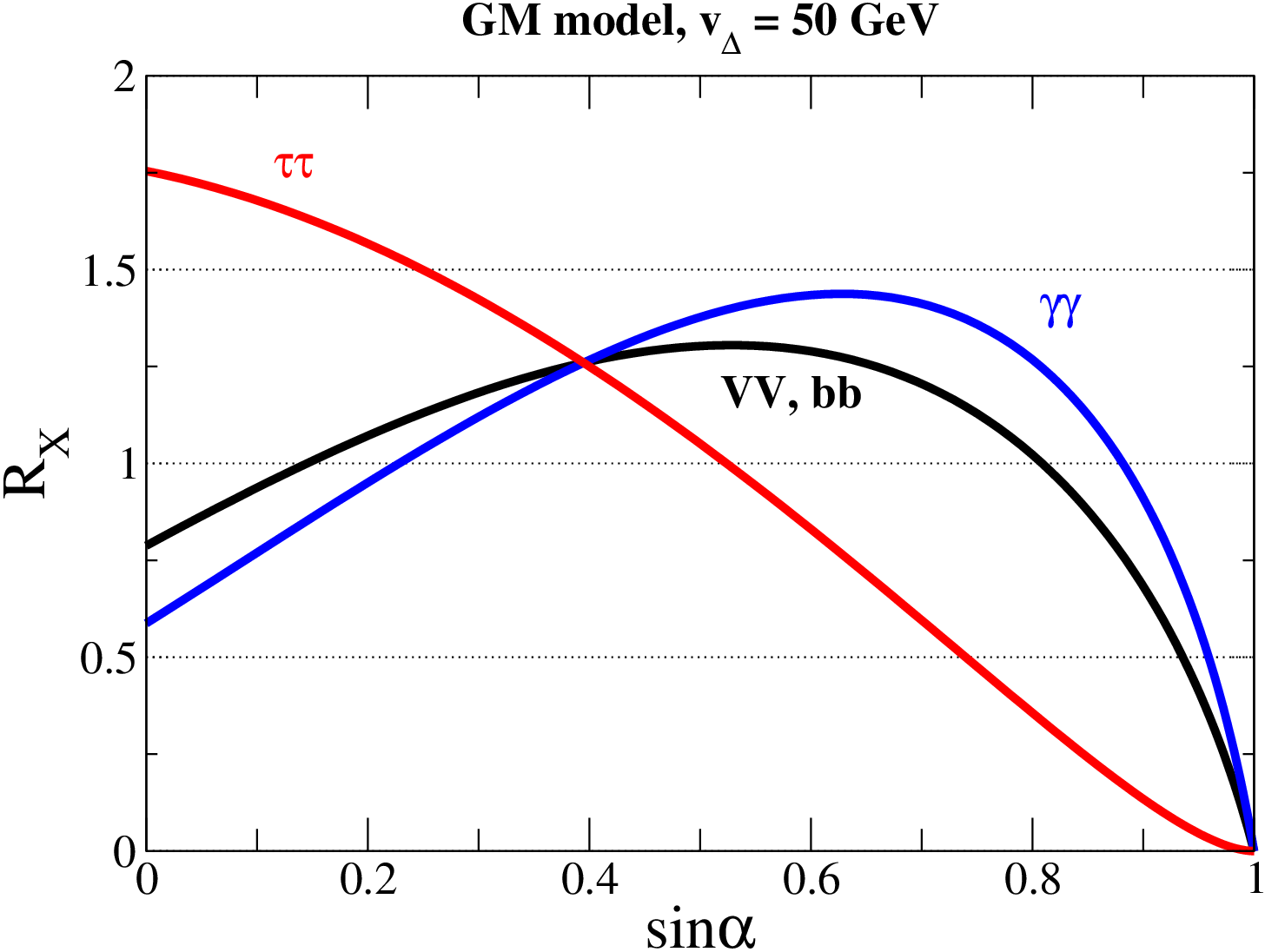}\hspace{3mm}
\includegraphics[width=50mm]{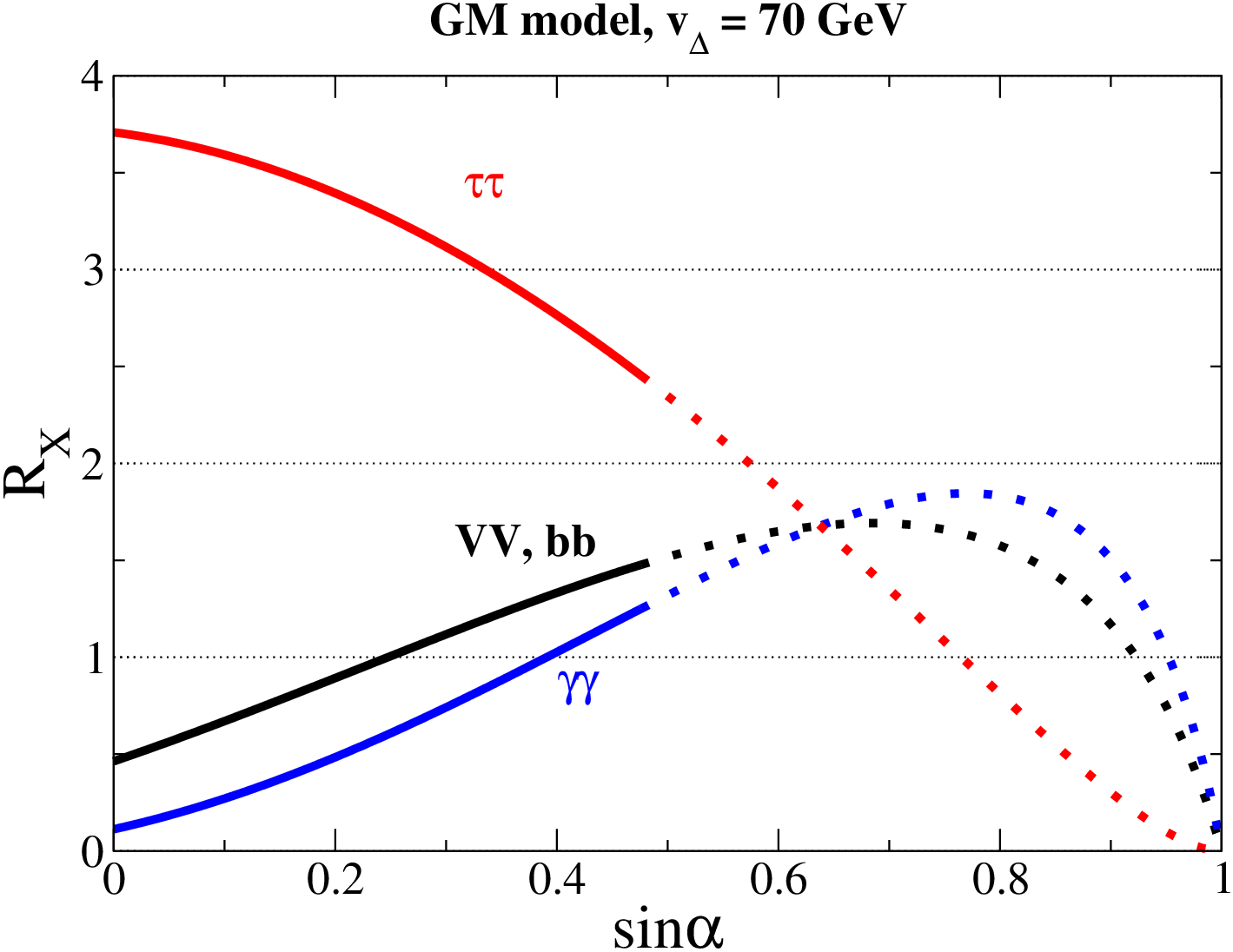}
\caption{$R_{X}$ for the various modes as a function of $\sin\alpha$ in the GM model. 
The left, center and right panels show the case of $v_\Delta = 30$ GeV, 50 GeV and 70 GeV, respectively. 
In the right panel, $R_X$ values indicated by dotted curves are excluded by the $S$ and $T$ parameters at the 95\% confidence level. }
\label{FIG:R2}
\end{center}
\end{figure}

In Fig.~\ref{FIG:R2}, the $R_X$ values are shown as a function of $\sin\alpha$ in the GM model 
for the cases of $v_\Delta=30$, 50 and 70 GeV. 
In the case of $v_\Delta=70$ GeV, the region of $\sin\alpha \gtrsim 0.48$ is excluded by the constraint from the $S$ parameter 
at the 95\% confidence level. 
Similar to the cases in the cHTM and rHTM, 
$R_{\tau\tau}$ is larger than 1 in the regions of small $\sin\alpha$. 
The $R_{VV}$ and $R_{b\bar{b}}$ values are larger than 1 when $\sin\alpha\gtrsim 0.07$, $0.15$ and 
0.24 in the cases of $v_\Delta=30$, 50 and 70 GeV, respectively. 
The $\sin\alpha$ dependence of $R_{\gamma\gamma}$ is similar to that for $R_{VV}$ and $R_{b\bar{b}}$, but the 
maximal allowed values of $R_{\gamma\gamma}$ are larger than those of $R_{VV}$ and $R_{b\bar{b}}$. 
The allowed maximal values for $R_{VV}$ and ($R_{\gamma\gamma}$) are about 
1.1 (1.2) for $v_\Delta=30$ GeV, 1.3 (1.4) for $v_\Delta=50$ GeV and 1.5 (1.3) for $v_\Delta=70$ GeV.

\begin{figure}[t]
\begin{center}
\includegraphics[width=50mm]{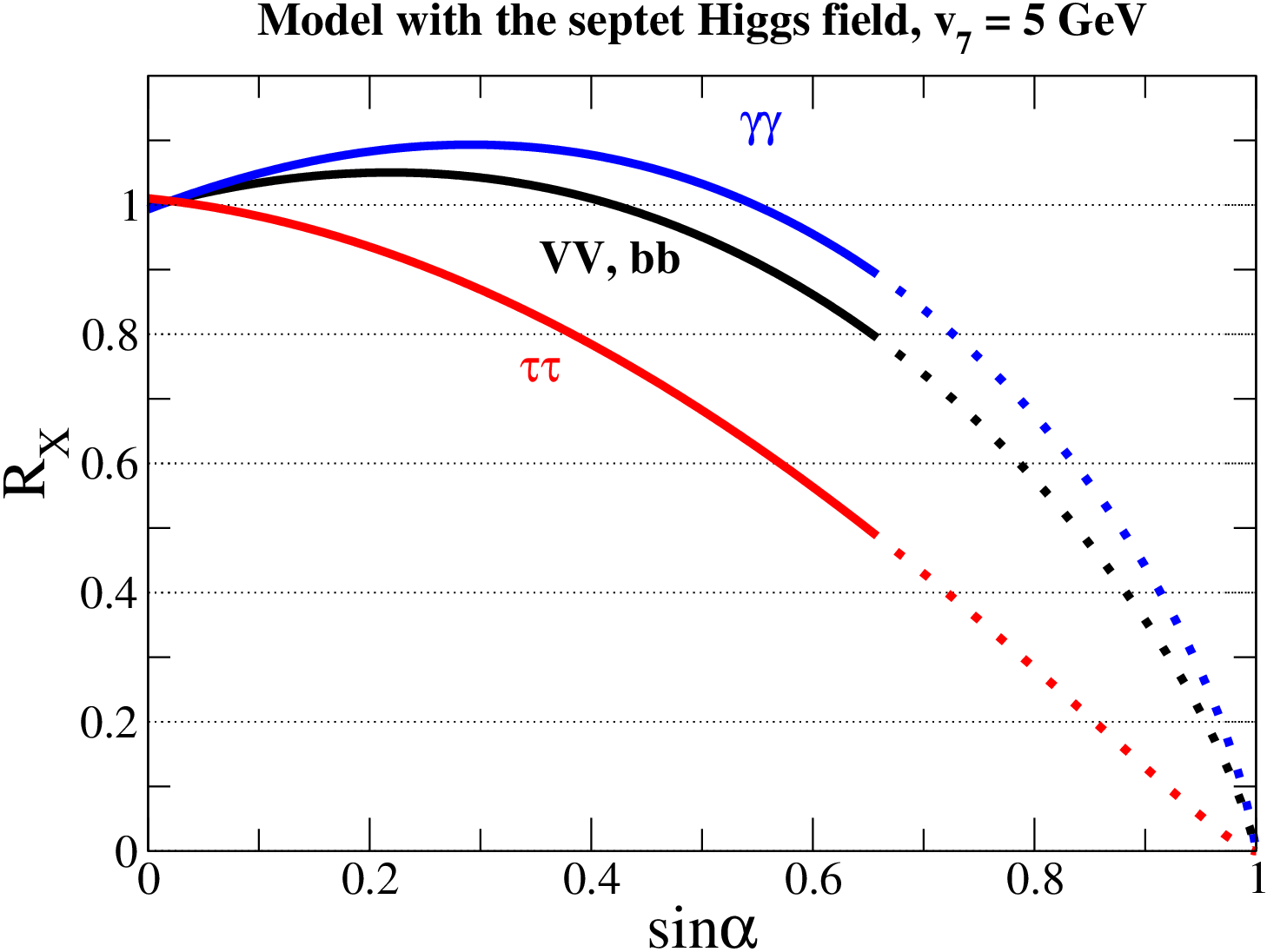}\hspace{3mm}
\includegraphics[width=50mm]{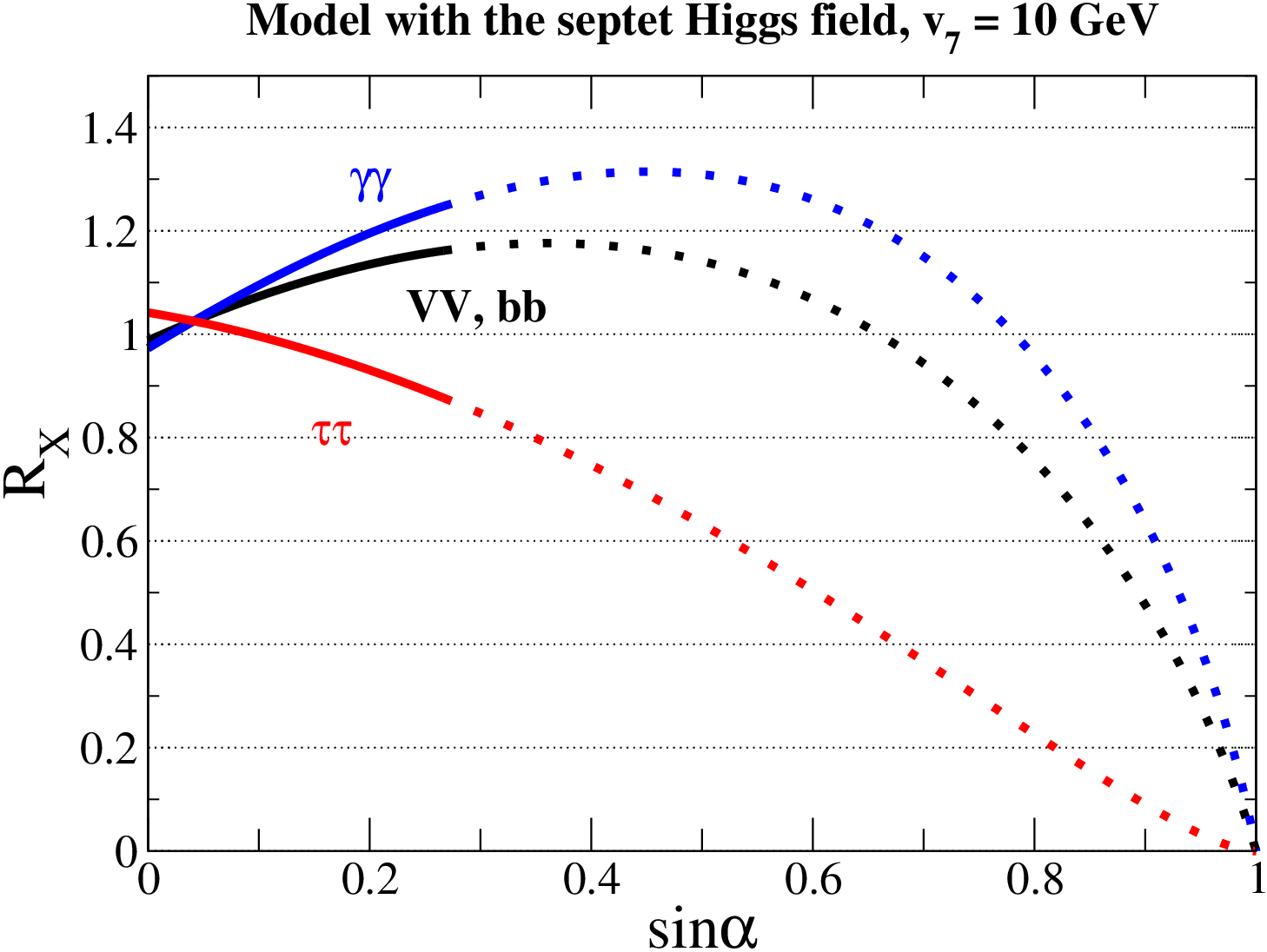}\hspace{3mm}
\includegraphics[width=50mm]{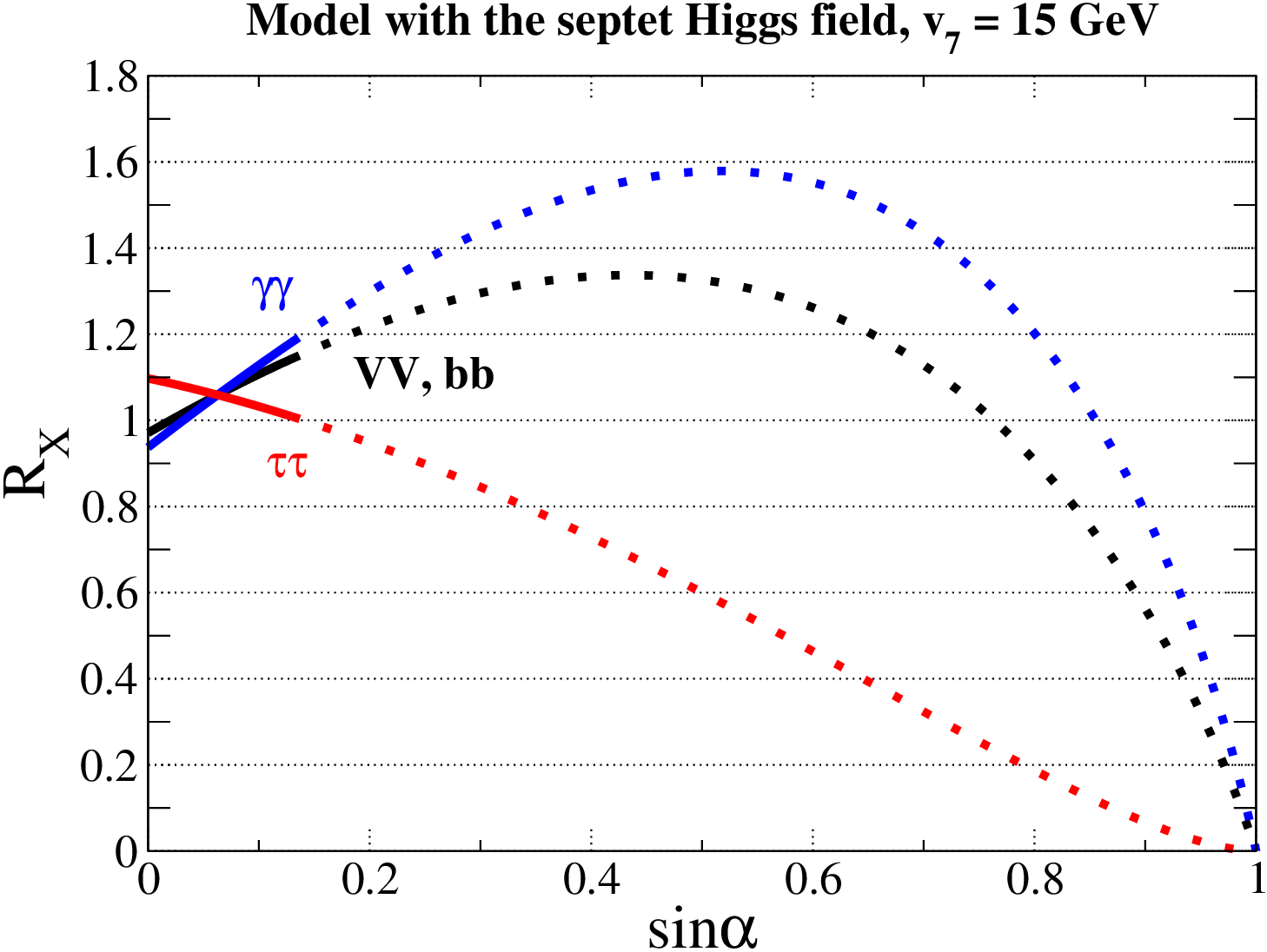}
\caption{$R_{X}$ for the various modes as a function of $\sin\alpha$ in the model with the septet Higgs field $\varphi_7$. 
The left, center and right panels show the cases of $v_\Delta = 5$ GeV, 10 GeV and 15 GeV, respectively. 
In all the figures, $R_X$ values indicated by dotted curves are excluded by the $S$ and $T$ parameters at the 95\% confidence level.}
\label{FIG:R3}
\end{center}
\end{figure}

Finally, we show the $\sin\alpha$ dependence in $R_X$ in the model with $\varphi_7$ in Fig.~\ref{FIG:R3}. 
We take the septet VEV $v_7$ to be 5 GeV (left panel), 10 GeV (center panel) and 15 GeV (right panel). 
The values of $R_X$ excluded by the constraint from the $S$ and $T$ parameters at the 
at the 95\% confidence level are shown as the dotted curves. 
The maximal allowed values of $R_{VV}$ ($R_{\gamma\gamma}$) are about 1.05 (1.10) for $v_7=5$ GeV, 
about 1.16 (1.25) for $v_7=10$ GeV and about 1.14 (1.20) for $v_7=15$ GeV. 

\section{Conclusions}

We have calculated the Higgs boson couplings with the gauge bosons 
$hZZ$ and $hWW$ as well as the fermions $hf\bar{f}$ in the general Higgs sector at the tree level. 
We have found that the $hZZ$ and $hWW$ couplings in Higgs sectors with exotic representation fields
can be larger than those in the SM. 
We also have studied the ratio of the event rates $R_X$ for $X=WW^*$, $ZZ^*$, $\gamma\gamma$, $b\bar{b}$ and $\tau\tau$ in 
various Higgs sectors. 
We have numerically evaluated the deviations in the Higgs boson couplings $c_{hVV}$ and $c_{hff}$ and the values for $R_X$ in the 
cHTM, the rHTM, the GM model and the model with the septet scalar field. 
We have found that the possible allowed magnitude of the deviation in the $hVV$ coupling 
can be as large as $\mathcal{O}(0.1)\%$ in the cHTM and rHTM, $\mathcal{O}(30)\%$ in the GM model and $\mathcal{O}(10)\%$ in 
the model with the septet field in the allowed parameter regions by the $S$ and $T$ parameters. 
By measuring the Higgs boson couplings precisely, we can get useful information to determine the structure of the 
Higgs sector. 
\\\\
\noindent
$Acknowledgments$

The authors would like to thank Lu-Hsing Tsai for useful discussions.
S.K. was supported in part by Grant-in-Aid for Scientific Research, Nos. 22244031, 
23104006 and 24340046.
K.Y. was supported in part by the National Science Council of R.O.C. under Grant No. NSC-101-2811-M-008-014.

\begin{appendix}
\section{Model with a septet Higgs field}

We discuss the Higgs sector with the doublet field $\phi$ and the septet field $\varphi_7$ with $Y=2$. 
These two Higgs fields can be expressed in the tensor form as $\phi_a$ and $(\varphi_7)_{ijklmn}$ 
where interchanges among the subscripts ($i,j,k,l,m,n$) are symmetric.
Component scalar fields can be specified as
\begin{align}
\phi_1 = \phi^+,~\phi_2=\phi^0=\frac{1}{\sqrt{2}}(h_\phi+v_\phi+iz_\phi), 
\end{align}
and
\begin{align}
&(\varphi_7)_{111111}=\varphi_7^{5+},~
(\varphi_7)_{211111}=\frac{\varphi_7^{4+}}{\sqrt{6}},~
(\varphi_7)_{221111}=\frac{\varphi_7^{3+}}{\sqrt{15}},~
(\varphi_7)_{222111}=\frac{\varphi_7^{++}}{\sqrt{20}},\notag\\
&(\varphi_7)_{222211}=\frac{\varphi_7^+}{\sqrt{15}},~
(\varphi_7)_{222221}=\frac{\varphi_7^0}{\sqrt{6}}=\frac{1}{\sqrt{12}}(h_7+v_7+iz_7),~
(\varphi_7)_{222222}=\bar{\varphi}_7^-. 
\end{align}
The most general Higgs potential is given by 
\begin{align}
V(\phi,\varphi_7)&=m^2|\phi|^2+m_7^2 |\varphi_7|^2+\lambda|\phi|^4+\lambda_1(|\varphi_7|^4)_1
+\lambda_2(|\varphi_7|^4)_2+\lambda_3(|\varphi_7|^4)_3+\lambda_4(|\varphi_7|^4)_4\notag\\
&+\kappa_1 (|\phi|^2|\varphi_7|^2)_1+\kappa_2 (|\phi|^2|\varphi_7|^2)_2, 
\end{align}
where $|\varphi_7|^2$ term can be expanded by 
\begin{align}
|\varphi_7|^2&=(\varphi_7^*)_{ijklmn}(\varphi_7)_{ijklmn}\notag\\
&=\varphi_7^{5+}\varphi_7^{5-}+\varphi_7^{4+}\varphi_7^{4-}
+\varphi_7^{3+}\varphi_7^{3-}+\varphi_7^{++}\varphi_7^{--}+\varphi_7^{+}\varphi_7^{-}
+\varphi_7^{0*}\varphi_7^{0}+\bar{\varphi}_7^{+}\bar{\varphi}_7^{-}. 
\end{align}
There are four independent $(|\varphi_7|^4)_\alpha$ ($\alpha=1,\dots,4$) terms, they can be explicitly written by 
\begin{align}
(|\varphi_7|^4)_1 &= (\varphi_7^*)_{ijklmn} (\varphi_7^*)_{abcdef} (\varphi_7)_{ijklmn}  (\varphi_7)_{abcdef}, \\
(|\varphi_7|^4)_2 &= (\varphi_7^*)_{ijklmn} (\varphi_7^*)_{abcdef} (\varphi_7)_{ijklmf}  (\varphi_7)_{abcden}, \\
(|\varphi_7|^4)_3 &= (\varphi_7^*)_{ijklmn} (\varphi_7^*)_{abcdef} (\varphi_7)_{ijklef}  (\varphi_7)_{abcdmn}, \\
(|\varphi_7|^4)_4 &= (\varphi_7^*)_{ijklmn} (\varphi_7^*)_{abcdef} (\varphi_7)_{ijkdef}  (\varphi_7)_{abclmn}. 
\end{align}
In addition to the $(|\varphi_7|^4)_\alpha$ terms, there are two independent $(|\phi|^2|\varphi|^2)_\beta$ ($\beta=1,2$) 
terms, they can be explicitly written by  
\begin{align}
(|\phi^2||\varphi_7|^2)_1 &= (\phi^*)_a(\phi)_a(\varphi_7^*)_{ijklmn} (\varphi_7)_{ijklmn} , \\
(|\phi^2||\varphi_7|^2)_2 &= (\phi^*)_a(\phi)_b(\varphi_7^*)_{ijklmb} (\varphi_7)_{ijklma}. 
\end{align}
The other combination $(\phi^*)_a(\epsilon)_{ab}(\varphi_7^*)_{ijklmb}(\phi)_c (\epsilon)_{cd}(\varphi_7)_{ijklmd}$ 
is the same as $(|\phi^2||\varphi_7|^2)_1-(|\phi^2||\varphi_7|^2)_2$. 

There is an accidental global $U(1)$ symmetry in the Higgs potential; i.e., 
the potential is invariant under the phase transformation of $\varphi_7$~\cite{Logan}. 
This $U(1)$ symmetry is spontaneously broken down due to the non-zero value of the VEV of $\varphi_7$. 
Thus, there appears a massless NG boson in addition to the usual NG bosons absorbed by the longitudinal components of $W$ and $Z$ bosons. 
There are several ways to avoid the appearance of the additional NG boson. 
For example, this NG boson can be absorbed by the additional neutral gauge boson by extending this global symmetry to 
the gauge symmetry via the Higgs mechanism. 
By introducing terms which break the $U(1)$ symmetry explicitly, we can also avoid such a massless scalar boson\footnote{
Quite recently, it has been considered to introduce higher dimensional operators to break the $U(1)$ symmetry explicitly\cite{Hisano_Tsumura}
in the Higgs sector with the septet. }.

From the tadpole condition, we obtain 
\begin{align}
m^2 &=-v_\varphi^2\lambda -\frac{v_7^2}{12} (6\kappa_1+5\kappa_2),\\
m_7^2 &= -\frac{v_\varphi^2 }{12}(6\kappa_1+5\kappa_2)-\frac{v_7^2}{18}(18\lambda_1+13\lambda_2+10\lambda_3+9\lambda_4). 
\end{align}
The mass matrix for the CP-even Higgs states in the basis of ($h_\phi,h_7$) is given by 
\begin{align}
M_{\text{CP-even}}^2 =
 \left(
\begin{array}{cc}
2v_\varphi^2\lambda & v_\varphi v_7 \left(\kappa_1+\frac{5}{6}\kappa_2\right)\\
v_\varphi v_7 \left(\kappa_1+\frac{5}{6}\kappa_2\right) & \frac{v_7^2}{9}(18\lambda_1+13\lambda_2+10\lambda_3+9\lambda_4)
\end{array}\right). 
\end{align}
That for the singly-charged Higgs states in the basis of ($\phi^\pm,\varphi_7^\pm,\bar{\varphi}_7^\pm$) is 
calculated by 
\begin{align}
M_+^2 =
 \left(
\begin{array}{ccc}
-\frac{v_7^2}{3}\kappa_2 & \frac{\sqrt{5}}{6\sqrt{2}}v_\varphi v_7 \kappa_2 & \frac{v_\varphi v_7}{2\sqrt{6}}\kappa_2 \\
\frac{\sqrt{5}}{6\sqrt{2}}v_\varphi v_7 \kappa_2 & -\frac{v_\varphi^2}{12}\kappa_2+\frac{v_7^2}{30}\bar{\lambda}& \frac{v_7^2}{6\sqrt{15}}\bar{\lambda} \\
\frac{v_\varphi v_7}{2\sqrt{6}}\kappa_2 & \frac{v_7^2}{6\sqrt{15}}\bar{\lambda} & \frac{v_\varphi^2}{12}\kappa_2+\frac{v_7^2}{18}\bar{\lambda}
\end{array}\right), 
\end{align}
where 
\begin{align}
\bar{\lambda}=5\lambda_2+8\lambda_3+9\lambda_4. 
\end{align}
The mass eigenstates of the CP-even Higgs bosons as well as the singly-charged Higgs bosons can be defined by introducing the following orthogonal matrix:
\begin{align}
\left(
\begin{array}{c}
h_\phi\\
h_7
\end{array}\right) =O_{\text{CP-even}}
\left(
\begin{array}{c}
h \\
H 
\end{array}\right),~
\left(
\begin{array}{c}
\phi^\pm\\
\varphi_7^\pm \\
\bar{\varphi}_7^\pm
\end{array}\right) =O_+
\left(
\begin{array}{c}
G^\pm \\
H^\pm \\
\bar{H}^\pm
\end{array}\right), 
\end{align}
where 
\begin{align}
&O_{\text{CP-even}} = 
\left(
\begin{array}{cc}
\cos\alpha & -\sin\alpha \\
\sin\alpha & \cos\alpha
\end{array}\right),  \label{alpha_7} \\
&O_+ = R_\theta R_{G^+}=
\left(
\begin{array}{ccc}
1 & 0 & 0 \\
0 & \cos\theta & -\sin\theta \\
0 & \sin\theta & \cos\theta
\end{array}\right)
\left(
\begin{array}{ccc}
-\frac{v_\phi}{\sqrt{v_\phi^2+16v_7^2}} & \frac{\sqrt{10}v_7}{\sqrt{v_\phi^2+10v_7^2}} & \frac{v_\phi}{\sqrt{v_\phi^2+16v_7^2}}\frac{\sqrt{6} v_7}{\sqrt{v_\phi^2+10v_7^2}} \\
-\frac{\sqrt{10}v_7}{\sqrt{v_\phi^2+16v_7^2}} &-\frac{v_\phi}{\sqrt{v_\phi^2+10v_7^2}}  & \frac{\sqrt{6}v_7}{\sqrt{v_\phi^2+16v_7^2}}\frac{\sqrt{10}v_7}{\sqrt{v_\phi^2+10v_7^2}} \\
\frac{\sqrt{6}v_7}{\sqrt{v_\phi^2+16v_7^2}} & 0 & \frac{\sqrt{v_\phi^2+10v_7^2}}{\sqrt{v_\phi^2+16v_7^2}}
\end{array}\right). \label{theta7}
\end{align}
The mass matrix is then transformed as
\begin{align}
O_+^T M_+^2 O_+ = R_\theta^T
\left(
\begin{array}{ccc}
0 & 0 & 0 \\
0 & (M_+^2)_{11} & (M_+^2)_{12} \\
0 & (M_+^2)_{12} & (M_+^2)_{22}
\end{array}\right)R_\theta
=\left(
\begin{array}{ccc}
0 & 0 & 0 \\
0 & m_{H^+}^2 & 0 \\
0 & 0 & m_{\bar{H}^+}^2
\end{array}\right), 
\end{align}
with
\begin{align}
(M_+^2)_{11} &= \frac{1}{v_\phi^2+10v_7^2}\left[-\frac{1}{12}\left(v_\phi^4+20v_\phi^2v_7^2+40v_7^4\right)\kappa_2+\frac{v_7^2v_\phi^2}{30}\bar{\lambda}\right],\\
(M_+^2)_{22} &= \frac{v_\phi^2+16v_7^2}{36(v_\phi^2+10v_7^2)}\left[3v_\phi^2\kappa_2+2v_7^2\bar{\lambda}\right], \\
(M_+^2)_{12} &= \frac{v_7^2v_\phi\sqrt{v_\phi^2+16v_7^2}}{6\sqrt{15}(v_\phi^2+10v_7^2)}
\left(15\kappa_2-\bar{\lambda}\right). 
\end{align}
The other multi-charged Higgs boson masses are calculated as
\begin{align}
m_{\varphi_7^{++}}^2 &= -\frac{1}{6}\left(\kappa_2v_\phi^2+\frac{4}{3}v_7^2\lambda_2\right)-\frac{4}{45}v_7^2\lambda_3,\\
m_{\varphi_7^{3+}}^2 &= -\frac{1}{4}\left(\kappa_2v_\phi^2+\frac{4}{3}v_7^2\lambda_2\right)-\frac{3v_7^2}{9}\lambda_3-\frac{3v_7^2}{10}\lambda_4,\\
m_{\varphi_7^{4+}}^2 &= -\frac{1}{3}\left(\kappa_2v_\phi^2+\frac{4}{3}v_7^2\lambda_2\right)-\frac{4v_7^2}{9}\lambda_3-\frac{v_7^2}{2}\lambda_4,\\
m_{\varphi_7^{5+}}^2 &= -\frac{5}{12}\left(\kappa_2v_\phi^2+\frac{4}{3}v_7^2\lambda_2\right)-\frac{5v_7^2}{9}\lambda_3-\frac{v_7^2}{2}\lambda_4.
\end{align}

\section{Gauge boson two point functions}

In this appendix, we give the analytic expressions for the 1PI diagram contribution to the gauge boson two point functions 
$\Pi_{XY}^{\text{1PI}}(p^2)$ in 
terms of the Passarino-Veltman functions~\cite{PV}, which are used to calculate the $S$ and $T$ parameters. 
Calculations are performed in the 't Hooft-Feynman gauge, so that the masses of the NG bosons $m_{G^+}$ and $m_{G^0}$ 
should be replaced by those corresponding gauge bosons; i.e., $m_{G^+}=m_W$ and $m_{G^0}=m_Z$. 
The following expressions are subtracted the contributions which appear in the SM. 

We first show the formulae in the GM model. 
In this model, when the Higgs potential is constructed under the custodial $SU(2)_V$ symmetric way, 
the physical Higgs bosons can be classified into the $SU(2)_V$ 5-plet ($H_5^{\pm\pm},H_5^\pm,H_5^0$), 
3-plet ($H_3^\pm,H_3^0$) and singlet Higgs bosons $H_1^0$ and $h$. 
The masses of the Higgs bosons belonging to the same $SU(2)_V$ multi-plet are the same; namely, 
$m_{H_5^{++}}=m_{H_5^+}=m_{H_5^0}$ and $m_{H_3^+}=m_{H_3^0}$. 
Detailed expressions for the relations between the mass eigenstates and the weak eigenstates of the scalar bosons and 
those masses are given in Refs.~\cite{Gunion-Vega-Wudka,AK_GM,Haber_Logan}. 

The $\Pi_{XY}^{\text{1PI}}(p^2)$ functions are then expressed by
\begin{align}
\Pi_{WW}^{\text{1PI}}(p^2)
&=\frac{g^2}{16\pi^2}\Big[
\frac{1}{2}B_5(p^2,m_{H_5^{++}},m_{H_5^{+}})
+\frac{c_\beta^2}{2}B_5(p^2,m_{H_5^{++}},m_{H_3^{+}})
+\frac{s_\beta^2}{2}B_5(p^2,m_{H_5^{++}},m_{G^{+}})\notag\\
&+\frac{3}{4}B_5(p^2,m_{H_5^{+}},m_{H_5^{0}}) 
+\frac{c_\beta^2}{4}B_5(p^2,m_{H_5^{+}},m_{H_3^{0}}) 
+\frac{s_\beta^2}{4}B_5(p^2,m_{H_5^{+}},m_{G^{0}}) \notag\\
&+\frac{c_\beta^2}{12}B_5(p^2,m_{H_3^{+}},m_{H_5^{0}}) +\frac{s_\beta^2}{12}B_5(p^2,m_{G^{+}},m_{H_5^{0}}) 
+\frac{1}{4}B_5(p^2,m_{H_3^{+}},m_{H_3^{0}})\notag\\
&+\frac{1}{4}\left(\frac{2}{3}\sqrt{6}c_\alpha c_\beta+s_\alpha s_\beta\right)^2B_5(p^2,m_{H_3^{+}},m_{H_1^0})
+\frac{1}{4}\left(-\frac{2}{3}\sqrt{6}s_\alpha c_\beta+c_\alpha s_\beta\right)^2B_5(p^2,m_{H_3^{+}},m_{h}) \notag\\
&+\frac{1}{4}\left(-\frac{2}{3}\sqrt{6}c_\alpha s_\beta+s_\alpha c_\beta\right)^2B_5(p^2,m_{G^{+}},m_{H_1^0})
+\frac{1}{4}\left[\Big(c_\alpha c_\beta+\frac{2}{3}\sqrt{6}s_\alpha s_\beta\Big)^2-1\right]B_5(p^2,m_{G^{+}},m_{h})\Big]\notag\\
&+\frac{g^2m_W^2}{16\pi^2}\Big[2s_\beta^2B_0(p^2,m_{H_5^{++}},m_W)+\frac{s_\beta^2}{c_W^2}B_0(p^2,m_{H_5^{+}},m_Z)
+\frac{s_\beta^2}{3}B_0(p^2,m_{H_5^{0}},m_W)\notag\\
&+\left(-s_\alpha c_\beta+\frac{2}{3}\sqrt{6}c_\alpha s_\beta\right)^2B_0(p^2,m_{H_1^{0}},m_W) 
+\left[\Big(c_\alpha c_\beta+\frac{2}{3}\sqrt{6}s_\alpha s_\beta\Big)^2-1\right]B_0(p^2,m_h,m_W) \Big],\\
\Pi_{ZZ}^{\text{1PI}}(p^2)
&=\frac{g_Z^2}{16\pi^2}\Big[c_{2W}^2B_5(p^2,m_{H_5^{++}},m_{H_5^{++}})
+\frac{c_{2W}^2}{4}B_5(p^2,m_{H_5^{+}},m_{H_5^{+}})
+\frac{c_{2W}^2}{4}B_5(p^2,m_{H_3^{+}},m_{H_3^{+}})\notag\\
&+\frac{c_\beta^2}{2}B_5(p^2,m_{H_5^{+}},m_{H_3^{+}})
+\frac{s_\beta^2}{2}B_5(p^2,m_{H_5^{+}},m_{G^{+}})\notag\\
&+\frac{c_\beta^2}{3}B_5(p^2,m_{H_5^{0}},m_{H_3^{0}})
+\frac{s_\beta^2}{3}B_5(p^2,m_{H_5^{0}},m_{G^{0}})\notag\\
&+\frac{1}{4}\left(\frac{2}{3}\sqrt{6}c_\alpha c_\beta+s_\alpha s_\beta\right)^2B_5(p^2,m_{H_3^{0}},m_{H_1^{0}})
+\frac{1}{4}\left(-\frac{2}{3}\sqrt{6}s_\alpha c_\beta+c_\alpha s_\beta\right)^2B_5(p^2,m_{H_3^{0}},m_h)\notag\\
&+\frac{1}{4}\left(-s_\alpha c_\beta+\frac{2}{3}\sqrt{6}c_\alpha s_\beta\right)^2B_5(p^2,m_{H_1^0},m_{G^{0}})
+\frac{1}{4}\left[\Big(c_\alpha c_\beta+\frac{2}{3}\sqrt{6}s_\alpha s_\beta\Big)^2-1\right]B_5(p^2,m_{h},m_{G^{0}})
\Big]\notag\\
&+\frac{g_Z^2m_Z^2}{16\pi^2}\Big[
\frac{4}{3}s_\beta^2B_0(p^2,m_{H_5^0},m_{Z})+2s_\beta^2c_W^2B_0(p^2,m_{H_5^+},m_W)\notag\\
&+\left(-s_\alpha c_\beta+\frac{2}{3}\sqrt{6}c_\alpha s_\beta\right)^2B_0(p^2,m_{H_1^{0}},m_Z) 
+\left[\Big(c_\alpha c_\beta+\frac{2}{3}\sqrt{6}s_\alpha s_\beta\Big)^2-1\right]B_0(p^2,m_h,m_Z)\Big],\\
\Pi_{\gamma\gamma}^{\text{1PI}}(p^2)
&=\frac{e^2}{16\pi^2}\Big[4B_5(p^2,m_{H_5^{++}},m_{H_5^{++}})
+B_5(p^2,m_{H_5^{+}},m_{H_5^{+}})
+B_5(p^2,m_{H_3^{+}},m_{H_3^{+}})\Big],\\
\Pi_{Z\gamma}^{\text{1PI}}(p^2)
&=\frac{eg_Z}{16\pi^2}\Big[2c_{2W}B_5(p^2,m_{H_5^{++}},m_{H_5^{++}})
+\frac{c_{2W}}{2}B_5(p^2,m_{H_5^{+}},m_{H_5^{+}})+\frac{c_{2W}}{2}B_5(p^2,m_{H_3^{+}},m_{H_3^{+}})
\Big], 
\end{align}
where $B_5(p^2,m_1,m_2)=A(m_1)+A(m_2)-4B_{22}(p^2,m_1,m_2)$~\cite{HHKM}. 

Next, $\Pi_{XY}^{\text{1PI}}(p^2)$ functions are calculated in model with the septet Higgs field in the case of $\theta=0$ defined in 
Eq.~(\ref{theta7}) as 
\begin{align}
\Pi_{WW}^{\text{1PI}}(p^2)
&=\frac{g^2}{16\pi^2}\Big[
3B_5(p^2,m_{\varphi_7^{5+}},m_{\varphi_7^{4+}})
+5B_5(p^2,m_{\varphi_7^{4+}},m_{\varphi_7^{3+}})
+6B_5(p^2,m_{\varphi_7^{3+}},m_{\varphi_7^{2+}})\notag\\
&+\frac{48c_\beta^2}{5+3c_\beta^2}B_5(p^2,m_{\varphi_7^{2+}},m_{H^+})
+\frac{45s_\beta^4}{4(5+3c_\beta^2)}B_5(p^2,m_{\varphi_7^{2+}},m_{\bar{H}^+})
+\frac{15s_\beta^2}{4}B_5(p^2,m_{\varphi_7^{2+}},m_{G^+})\notag\\
&+\frac{5(-4s_\alpha c_\beta+c_\alpha s_\beta)^2}{4(5+3c_\beta^2)}B_5(p^2,m_{H^+},m_{h})
+\frac{5(4c_\alpha c_\beta+s_\alpha s_\beta)^2}{4(5+3c_\beta^2)}B_5(p^2,m_{H^+},m_{H})\notag\\
&+\frac{5(5+3c_{2\beta})^2}{16(5+3c_\beta^2)}B_5(p^2,m_{H^+},m_{A})
+\frac{45s_\beta^2c_\beta^2}{4(5+3c_\beta^2)}B_5(p^2,m_{H^+},m_{G^0})\notag\\
&+\frac{3c_\beta^2(4s_\alpha c_\beta-c_\alpha s_\beta)^2}{4(5+3c_\beta^2)}B_5(p^2,m_{\bar{H}^+},m_{h})
+\frac{3c_\beta^2(4c_\alpha c_\beta+s_\alpha s_\beta)^2}{4(5+3c_\beta^2)}B_5(p^2,m_{\bar{H}^+},m_{H})\notag\\
&+\frac{12c_\beta^2}{5+3c_\beta^2}B_5(p^2,m_{\bar{H}^+},m_{A})
+\frac{75s_\beta^2}{4(5+3c_\beta^2)}B_5(p^2,m_{\bar{H}^+},m_{G^0})\notag\\
&+\frac{1}{4}(s_\alpha c_\beta-4c_\alpha s_\beta)^2B_5(p^2,m_{G^+},m_{H})
+\frac{1}{4}\left[(c_\alpha c_\beta+4s_\alpha s_\beta)^2-1\right]B_5(p^2,m_{G^+},m_{h})
\Big]\notag\\
&+\frac{g^2m_W^2}{16\pi^2}\Big[15s_\beta^2B_0(p^2,m_{\varphi_7^{2+}},m_W)
+\frac{45c_\beta^2s_\beta^2}{(5+3c^2_\beta)c_W^2}B_0(p^2,m_{H^+},m_Z)\notag\\
&+\frac{75s_\beta^2}{(5+3c_\beta^2)c_W^2}B_0(p^2,m_{\bar{H}^+},m_Z)
\notag\\
&+(-s_\alpha c_\beta+4c_\alpha s_\beta)^2B_0(p^2,m_{H},m_W) 
+\left[(c_\alpha c_\beta+4s_\alpha s_\beta)^2-1\right]B_0(p^2,m_h,m_W)\Big],\\
\Pi_{ZZ}^{\text{1PI}}(p^2)
&=\frac{g_Z^2}{16\pi^2}\Big[(5c_W^2-2)^2B_5(p^2,m_{\varphi_7^{5+}},m_{\varphi_7^{5+}})
+(4c_W^2-2)^2B_5(p^2,m_{\varphi_7^{4+}},m_{\varphi_7^{4+}})\notag\\
&+(3c_W^2-2)^2B_5(p^2,m_{\varphi_7^{3+}},m_{\varphi_7^{3+}})
+(2c_W^2-2)^2B_5(p^2,m_{\varphi_7^{2+}},m_{\varphi_7^{2+}})\notag\\
&+\left[\frac{c_{2W}(3c_\beta^2+5)-24c_\beta^2}{10+6c_\beta^2}\right]^2B_5(p^2,m_{H^+},m_{H^+})\notag\\
&+\left(c_W^2+\frac{9}{2}-\frac{20}{5+3c_\beta^2}\right)^2B_5(p^2,m_{\bar{H}^+},m_{\bar{H}^+})
+\frac{135s_\beta^4c_\beta^2}{2(5+3c_\beta^2)^2}B_5(p^2,m_{H^+},m_{\bar{H}^+})\notag\\
&+\frac{45s_\beta^2c_\beta^2}{2(5+3c_\beta^2)}B_5(p^2,m_{H^+},m_{G^+})
+\frac{75s^2_\beta}{2(5+3c_\beta^2)}B_5(p^2,m_{\bar{H}^+},m_{G^+})
\notag\\
&+\frac{1}{4}(-4s_\alpha c_\beta  +c_\alpha s_\beta)^2B_5(p^2,m_{A},m_{h})
+\frac{1}{4}(4c_\alpha c_\beta  +s_\alpha s_\beta)^2B_5(p^2,m_{A},m_{H})\notag\\
&+\frac{1}{4}(s_\alpha c_\beta  -4c_\alpha s_\beta)^2B_5(p^2,m_{G^0},m_{H})
+\frac{1}{4}\left[(c_\alpha c_\beta  +4s_\alpha s_\beta)^2-1\right]B_5(p^2,m_{G^0},m_{h})\Big]\notag\\
&+\frac{g_Z^2m_Z^2}{16\pi^2}\Big[
\frac{90c_\beta^2s_\beta^2c_W^2}{5+3c^2_\beta}B_0(p^2,m_{H^+},m_W)
+\frac{150s_\beta^2c_W^2}{5+3c_\beta^2}B_0(p^2,m_{\bar{H}^+},m_W)
\notag\\
&+\left(-s_\alpha c_\beta+4c_\alpha s_\beta\right)^2B_0(p^2,m_{H},m_Z) 
+\left[(c_\alpha c_\beta+4s_\alpha s_\beta)^2-1\right]B_0(p^2,m_h,m_Z)\Big],
\end{align}
\begin{align}
\Pi_{\gamma\gamma}^{\text{1PI}}(p^2)
&=\frac{e^2}{16\pi^2}\Big[25B_5(p^2,m_{\varphi_7^{5+}},m_{\varphi_7^{5+}})
+16B_5(p^2,m_{\varphi_7^{4+}},m_{\varphi_7^{4+}})
+9B_5(p^2,m_{\varphi_7^{3+}},m_{\varphi_7^{3+}})\notag\\
&+4B_5(p^2,m_{\varphi_7^{++}},m_{\varphi_7^{++}})+B_5(p^2,m_{H^{+}},m_{H^{+}})
+B_5(p^2,m_{\bar{H}^{+}},m_{\bar{H}^{+}})\Big],\\
\Pi_{Z\gamma}^{\text{1PI}}(p^2)
&=\frac{eg_Z}{16\pi^2}\Big[5(5c_W^2-2)B_5(p^2,m_{\varphi_7^{5+}},m_{\varphi_7^{5+}})
+4(4c_W^2-2)B_5(p^2,m_{\varphi_7^{4+}},m_{\varphi_7^{4+}})\notag\\
&+3(3c_W^2-2)B_5(p^2,m_{\varphi_7^{3+}},m_{\varphi_7^{3+}})
+2(2c_W^2-2)B_5(p^2,m_{\varphi_7^{++}},m_{\varphi_7^{++}})\notag\\
&+\frac{c_{2W}(3c_\beta^2+5)-24c_\beta^2}{10+6c_\beta^2}B_5(p^2,m_{H^{+}},m_{H^{+}})
+\left(c_W^2+\frac{9}{2}-\frac{20}{5+3c_\beta^2}\right)B_5(p^2,m_{\bar{H}^{+}},m_{\bar{H}^{+}})
\Big].
\end{align}

\end{appendix}

\end{document}